\documentclass[a4paper,english,pra,twocolumn]{revtex4-1}
\usepackage{libertine-type1}
\usepackage[T1]{fontenc}
\usepackage[latin9]{inputenc}
\usepackage{amsthm}
\usepackage{amsmath}
\usepackage{amssymb}
\usepackage{setspace}

\makeatletter

\providecommand{\tabularnewline}{\\}

\@ifundefined{textcolor}{}
{%
 \definecolor{BLACK}{gray}{0}
 \definecolor{WHITE}{gray}{1}
 \definecolor{RED}{rgb}{1,0,0}
 \definecolor{GREEN}{rgb}{0,1,0}
 \definecolor{BLUE}{rgb}{0,0,1}
 \definecolor{CYAN}{cmyk}{1,0,0,0}
 \definecolor{MAGENTA}{cmyk}{0,1,0,0}
 \definecolor{YELLOW}{cmyk}{0,0,1,0}
}
  \theoremstyle{definition}
  \newtheorem{defn}{\protect\definitionname}
  \theoremstyle{plain}
  \newtheorem{prop}{\protect\propositionname}
  \theoremstyle{plain}
  \newtheorem{lem}{\protect\lemmaname}
\theoremstyle{plain}
\newtheorem{thm}{\protect\theoremname}

\makeatother

\usepackage{babel}
  \providecommand{\definitionname}{Definition}
  \providecommand{\lemmaname}{Lemma}
  \providecommand{\propositionname}{Proposition}
\providecommand{\theoremname}{Theorem}

\begin{document}

\title{Strong Quantum Nonlocality without Entanglement }

\author{Saronath Halder}

\email{saronath.halder@gmail.com}

\affiliation{Department of Mathematics, Indian Institute of Science Education
and Research Berhampur, Transit Campus, Government ITI, Berhampur
760010, Odisha, India}

\author{Manik Banik}

\email{manik11ju@gmail.com}

\affiliation{S. N. Bose National Center for Basic Sciences, Block JD, Sector III,
Bidhannagar, Kolkata 700098, India}

\author{Sristy Agrawal}

\email{sristyagrawal8@gmail.com}

\affiliation{Indian Institute of Science Education and Research Kolkata, Mohanpur,
West Bengal 741246, India}

\author{Somshubhro Bandyopadhyay }

\email{som@jcbose.ac.in}

\affiliation{Department of Physics and Center for Astroparticle Physics and Space
Science, Bose Institute, EN 80, Sector V, Bidhannagar, Kolkata 700091,
India}
\begin{abstract}
Quantum nonlocality is usually associated with entangled states by
their violations of Bell-type inequalities. However, even unentangled
systems, whose parts may have been prepared separately, can show nonlocal
properties. In particular, a set of product states is said to exhibit
``quantum nonlocality without entanglement'' if the states are locally
indistinguishable; i.e., it is not possible to optimally distinguish
the states by any sequence of local operations and classical communication.
Here, we present a stronger manifestation of this kind of nonlocality
in multiparty systems through the notion of local irreducibility.
A set of multiparty orthogonal quantum states is defined to be locally
irreducible if it is not possible to locally eliminate one or more
states from the set while preserving orthogonality of the postmeasurement
states. Such a set, by definition, is locally indistinguishable, but
we show that the converse does not always hold. We provide the first
examples of orthogonal product bases on $\mathbb{C}^{d}\otimes\mathbb{C}^{d}\otimes\mathbb{C}^{d}$
for $d=3,4$ that are locally irreducible in all bipartitions, where
the construction for $d=3$ achieves the minimum dimension necessary
for such product states to exist. The existence of such product bases
implies that local implementation of a multiparty separable measurement
may require entangled resources across all bipartitions. 
\end{abstract}
\maketitle
Composite quantum systems, parts of which are physically separated,
can possess nonlocal properties. The most well-known manifestation
of quantum nonlocality---Bell nonlocality \cite{Brunner-et-al-2014},
arises from entangled states \cite{Entanglement-horodecki}. Entangled
states are nonlocal for they violate Bell-type inequalities \cite{Bell-1964,CHSH-1969,Freedman-Causer-1972,Aspect+-1981,Aspect+1982,Hensen+2015,handsteiner+2017,Rosenfeld+,BIG-BELL}---the
family of inequalities that must be satisfied by probabilities arising
from any local realistic theory. Apart from the foundational implications,
Bell nonlocality tests have applications in quantum technologies as
they quantify nonclassicality in a device-independent manner \cite{Barrett+2005,Acin+2006,Brunner+2008,Pironio+2010,Colbeck-Renner2012}.

Nonlocal properties, however, are not restricted only to entangled
systems. In the seminal paper \cite{ben99} \emph{Quantum nonlocality
without entanglement}, Bennett \emph{et al.} showed that product states
can exhibit nonlocal properties in a way fundamentally different from
Bell nonlocality. In particular, they considered the following problem:
Suppose that a quantum system, consisting of two parts held by separated
observers, was prepared in one of several known orthogonal product
states. The task is to identify, as well as possible, in which state
the system is in, using local operations and classical communication
(LOCC). The question they asked was whether for any known set of orthogonal
product states exact discrimination is always possible using LOCC.

Now intuition suggests that the answer to the above question ought
to be yes because product states admit local preparation (following
some known set of rules), and therefore, it should be possible to
learn about the state of the system with local measurements alone.
But surprisingly, the authors presented an orthogonal product basis
(OPB) on $\mathbb{C}^{3}\otimes\mathbb{C}^{3}$ for which exact discrimination
is not possible using LOCC \cite{ben99}. Subsequently, more such
examples were found in both bipartite \cite{ben99u,grois01,divin03,feng09,Yang15,xu16-1}
and multiparty systems \cite{ben99u,divin03,feng09,Yang13,nis06,halder,Xu-16-2,Xu-17,Wang-2017-Qinfoprocess,Zhang-Oh-2017}
and their properties explored \cite{ben99u,grois01,Cohen-2008,childs13,Corke17,MD-RA-2017}. 

The possibility of this kind of result was, in fact, first pointed
out by Peres and Wootters a few years earlier \cite{Peres-Wootters-1991}
(also see Refs. \cite{Massar-Popescu-1995,Wootters-2006}). For a
specific set of three nonorthogonal product states, they conjectured
that LOCC measurements are suboptimal for state discrimination. The
conjecture was only recently shown to be true \cite{Chitambar-2013}.
We now say that any set of product states that cannot be optimally
(exactly, if and only if the states are orthogonal) distinguished
by LOCC exhibit nonlocality without entanglement. Here, nonlocality
is in the sense that a measurement on the whole system reveals more
information about the state of the system than any sequence of LOCC
on their parts, even though they may have been prepared in different
labs. Let us also note that the recent PBR theorem \cite{Pusey+2012}
(also see Refs. \cite{Bandyopadhyay+2014,Mansfield-2016}) reveals
yet another nonlocal feature of nonorthogonal product states, where
state elimination with certainty becomes possible only by entangled
measurements on the joint system.

The results of Peres-Wootters \cite{Peres-Wootters-1991} and Bennett
\emph{et al.} \cite{ben99} initiated a plethora of studies on more
general local state discrimination problems---the task of optimal
discrimination of multiparty states, not necessarily product, by means
of LOCC \cite{ben99,ben99u,rin04,Wootters-2006,Chitambar-2013,Walgate-2000,Virmani-2001,Ghosh-2001,walgate-2002,HSSH,Ghosh-2004,Nathanson-2005,Watrous-2005,Hayashi-etal-2006,Duan2007,Calsamiglia-2010,fan-2005,feng09,BGK-2011,Bandyo-2011,Yu-Duan-2012,Nathanson-2010,Cosentino-Russo-2014,B-IQC-2015,Cosentino-2013,BN-2013,childs13,Duan-2009,Yang13,zhang14,zhang15,wang15,halder,Yang15,yu15,Xu-16-2,xu16-1,zhang16-1,Xu-17,Wang-2017-Qinfoprocess,wang17,zhang16,zhang17-1,zhang17-3,Zhang-Oh-2017,Corke17,MD-RA-2017}.
It was found that in some cases, e.g., a set of two pure states, LOCC
can indeed accomplish the task as efficiently as global measurements
\cite{Walgate-2000,Virmani-2001}, whereas in some other cases, e.g.,
orthogonal entangled bases \cite{Ghosh-2001,HSSH,Ghosh-2004,Nathanson-2005,fan-2005,Duan2007,BGK-2011,Cosentino-2013,Cosentino-Russo-2014,Yu-Duan-2012}
they cannot, and we call such states locally indistinguishable. Locally
indistinguishable states have found useful applications in quantum
cryptography primitives such as data hiding \cite{Terhal2001,DiVincenzo2002,Eggeling2002,MatthewsWehnerWinter}
and quantum secret sharing \cite{Markham-Sanders-2008}.

In this paper, we report new nonlocal properties of multiparty orthogonal
product states---within the framework of local state discrimination,
but considering instead a more basic problem---quantum state elimination
using orthogonality-preserving local measurements (a measurement is
orthogonality preserving if the the post-measurement states remain
orthogonal). The motivation stemmed from the observation that some
sets of orthogonal states on a composite Hilbert space are locally
reducible;\emph{ }i.e., it is possible to locally eliminate one or
more states from the set while preserving orthogonality of the postmeasurement
states. For such sets, the task of local state discrimination is therefore
reduced to that of a subset of states. 

While a locally distinguishable set is locally reducible (trivially),
the opposite is not true in general. In the following examples, the
locally indistinguishable sets are locally reducible to a union of
two or more disjoint subsets, each of which can be addressed individually.

(a) Consider the entangled orthogonal basis on $\mathbb{C}^{2}\otimes\mathbb{C}^{4}$:
\begin{equation}
\begin{array}{cccc}
\left|00\right\rangle \pm\left|11\right\rangle  &  &  & \left|02\right\rangle \pm\left|13\right\rangle \\
\left|01\right\rangle \pm\left|10\right\rangle  &  &  & \left|03\right\rangle \pm\left|12\right\rangle .
\end{array}\label{example-locally-reducible (a)}
\end{equation}
Here, Bob performs a local measurement to distinguish the subspaces
spanned by $\left\{ \left|0\right\rangle ,\left|1\right\rangle \right\} $
and $\left\{ \left|2\right\rangle ,\left|3\right\rangle \right\} $.
Depending upon the outcome, Alice and Bob end up with a state belonging
to one of the two subsets (left or right). Note that, neither subset
is locally distinguishable \cite{Ghosh-2001} (in fact, neither is
locally reducible---see Proposition \ref{bell-example}).

(b) Consider the orthogonal basis on $\mathbb{C}^{3}\otimes\mathbb{C}^{3}$:
\begin{equation}
\begin{array}{ccccccc}
\left|00\right\rangle \pm\left|11\right\rangle  &  & \left|02\right\rangle  &  & \left|20\right\rangle  &  & \left|22\right\rangle \\
\left|01\right\rangle \pm\left|10\right\rangle  &  & \left|12\right\rangle  &  & \left|21\right\rangle .
\end{array}\label{locally-reducible (b)}
\end{equation}
Here, if the unknown state is one of the product states, it can always
be correctly identified, and if it is not, all product states can
be locally eliminated. In the latter case, Alice and Bob will end
up with one of the four Bell states (local protocol given in Appendix
A). Note that, unlike the previous example, here not all the subsets
are locally indistinguishable. 

The above examples give rise to the following question: Are all locally
indistinguishable sets locally reducible? The answer is no. As will
be shown, some of the well-known locally indistinguishable sets are
not locally reducible. First, we have the following definition. 
\begin{defn}
(Locally irreducible set) A set of orthogonal quantum states on $\mathcal{H}=\bigotimes_{i=1}^{n}\mathcal{H}_{i}$
with $n\geq2$ and $\dim\mathcal{H}_{i}\geq2$, $i=1,\dots,n$, is
locally irreducible if it is not possible to eliminate one or more
states from the set by orthogonality-preserving local measurements. 
\end{defn}
A locally indistinguishable set in general is not locally irreducible
except when it contains three orthogonal pure states. 
\begin{prop}
\label{three-states-result} Any set of three locally indistinguishable
orthogonal pure states on $\mathcal{H}=\bigotimes_{i=1}^{n}\mathcal{H}_{i}$
with $n\geq2$ and $\dim\mathcal{H}_{i}\geq2$, $i=1,\dots,n$, is
locally irreducible. 
\end{prop}
Since any two orthogonal pure states can be exactly distinguished
by LOCC \cite{Walgate-2000}, a locally reducible set containing three
orthogonal pure states must be locally distinguishable. But this contradicts
the fact that the set is known to be locally indistinguishable. This
proves the proposition.

We will now describe a sufficient condition for local irreducibility.
The formalism was originally developed \cite{walgate-2002} (also
see Refs. \cite{zhang14,zhang15}) for local indistinguishability.
We begin by defining a nontrivial measurement \cite{walgate-2002}. 
\begin{defn}
A measurement is nontrivial if not all the POVM elements are proportional
to the identity operator. Otherwise, the measurement is trivial. 
\end{defn}
The crux of the argument \cite{walgate-2002} was that, in any local
protocol one of the parties must go first, and whoever goes first
must be able to perform some nontrivial orthogonality-preserving measurement
(NOPM). This fits naturally into our scenario for the following reasons.
The measurement should be orthogonality preserving because we require
that any measurement outcome must leave the postmeasurement states
mutually orthogonal, possibly eliminating some states but not all
(unless it correctly identifies the input right away). It is also
essential that the measurement is nontrivial because a trivial measurement
despite satisfying (trivially) the orthogonality-preserving conditions,
gives us no information about the state. The sufficient condition
follows by noting that, if none of the parties can perform a local
NOPM, the states must be locally irreducible.

Following Ref. \cite{walgate-2002} we now discuss how to apply this
condition when a set contains only orthogonal pure states. The basic
idea is to check whether an orthogonality-preserving POVM on any of
the subsystems is trivial or not. If it is trivial for all subsystems,
the states are locally irreducible\@.

Let $S=\left\{ \left|\psi_{i}\right\rangle \right\} _{i=1}^{k}$ be
a set of orthogonal pure states on $\mathcal{H}=\bigotimes_{i=1}^{n}\mathcal{H}_{i}$,
where $n\geq2$, and $\dim\mathcal{H}_{i}\geq2$, $i=1,\dots,n$.
Consider a POVM $\left\{ \pi_{\alpha}^{i}\right\} $, $\alpha=1,2,\dots$
that may be performed on the \emph{i}th subsystem. The POVM elements
$\pi_{\alpha}^{i}$ are positive operators summing up to identity
and correspond to the measurement outcomes. Further, each element
admits the Krauss form: $\pi_{\alpha}^{i}=M_{\alpha}^{i\dagger}M_{\alpha}^{i}$,
where $M_{\alpha}^{i}$s are the Krauss operators. The probability
that an input state $\left|\psi_{x}\right\rangle \in S$ yields the
outcome $\alpha$ is $p_{\alpha}=\left\langle \psi_{x}\left|\mathbb{I}_{1}\otimes\cdots\otimes\pi_{\alpha}^{i}\otimes\cdots\otimes\mathbb{I}_{n}\right|\psi_{x}\right\rangle $
with the corresponding postmeasurement state given by $\frac{1}{\sqrt{p_{\alpha}}}\left(\mathbb{I}_{1}\otimes\cdots\otimes M_{\alpha}^{i}\otimes\cdots\otimes\mathbb{I}_{n}\right)\left|\psi_{x}\right\rangle $.
Since we require the POVM to be orthogonality preserving, for all
pairs of states $\left\{ \left|\psi_{x}\right\rangle ,\left|\psi_{y}\right\rangle \right\} $
, $x\neq y$ and all outcomes $\alpha$, the conditions 
\begin{eqnarray}
\left\langle \psi_{x}\left|\mathbb{I}_{1}\otimes\cdots\otimes\pi_{\alpha}^{i}\otimes\cdots\otimes\mathbb{I}_{n}\right|\psi_{y}\right\rangle  & = & 0\label{IP}
\end{eqnarray}
need to be satisfied. To use the above conditions effectively, we
represent each POVM element $\pi_{\alpha}^{i}$, $\alpha=1,2,\dots$
by a $d_{i}\times d_{i}$ matrix (in the computational basis) and
solve for the matrix elements by choosing suitable pairs of vectors
(also expressed in the computational basis of $\mathcal{H}$). This
can be done exactly in many problems of interest. Now if we find that
the conditions (\ref{IP}) are satisfied only if $\pi_{\alpha}^{i}$
is proportional to the identity for all $\alpha$, then the measurement
is trivial. This means the $i^{{\rm th}}$ party cannot begin a LOCC
protocol, and if this is true for all $i$, then none of the parties
can go first. Therefore, $S$ is locally irreducible. We will use
this condition extensively in our proofs.

The OPB on $\mathbb{C}^{3}\otimes\mathbb{C}^{3}$ \cite{ben99} is
locally irreducible. This follows from the proof showing that the
states are locally indistinguishable \cite{walgate-2002}. We now
show that the Bell basis and the three-qubit GHZ basis are locally
irreducible (both are locally indistinguishable \cite{Ghosh-2001,HSSH})
using the method just described. 
\begin{prop}
\label{bell-example} The two-qubit Bell basis (unnormalized): $\left|00\right\rangle \pm\left|11\right\rangle ,\left|01\right\rangle \pm\left|10\right\rangle $
is locally irreducible. 
\end{prop}
The proof is by contradiction. Suppose that the Bell basis is locally
reducible. Then, either Alice or Bob must be able to begin the protocol
by performing some local NOPM. Without loss of generality assume that
Bob goes first. Bob's general measurement can be represented by a
set of $2\times2$ POVM elements $\pi_{\alpha}=\left(\begin{array}{cc}
a_{00} & a_{01}\\
a_{10} & a_{11}
\end{array}\right)$ written in the $\left\{ \left|0\right\rangle ,\left|1\right\rangle \right\} $
basis. Since this measurement is orthogonality preserving, for any
pair of Bell states the conditions (\ref{IP}) must hold. By choosing
suitable pairs, it is easy to show that $\pi_{\alpha}$ must be proportional
to the identity (details in Appendix B). As the argument holds for
all outcomes, all of Bob's POVM elements are proportional to the identity.
This means Bob cannot go first, and from the symmetry of the Bell
states, neither can Alice. This completes the proof. 
\begin{prop}
\label{2X2X2-GHZ} The three-qubit GHZ basis (unnormalized): $\left|000\right\rangle \pm\left|111\right\rangle $,
$\left|011\right\rangle \pm\left|100\right\rangle $, $\left|001\right\rangle \pm\left|110\right\rangle $,
$\left|010\right\rangle \pm\left|101\right\rangle $, is locally irreducible. 
\end{prop}
The proof is along the same lines as in the previous one and is given
in Appendix C (can be extended for a $N$-qubit GHZ basis).

We now come to the main part of the paper. Here we consider the following
question: Do there exist multiparty orthogonal sets that are locally
irreducible in every bipartition? The motivation for asking this question
is that many of the properties of multiparty states in general are
not preserved if we change the spatial configuration. For example,
the three-party ($A$, $B$, and $C$) unextendible product basis
(UPB) on $\mathbb{C}^{2}\otimes\mathbb{C}^{2}\otimes\mathbb{C}^{2}$
\cite{ben99u} is locally indistinguishable (hence, nonlocal when
all parts are separated) but can be perfectly distinguished across
all bipartitions $A\vert BC$, $B\vert CA$, and $C\vert AB$ \cite{ben99u}
using LOCC (and therefore, not nonlocal in the bipartitions). In fact,
one can also find sets of entangled states that are locally distinguishable
in one bipartition but not in others (see Appendix D).

So which sets of orthogonal states are expected to remain locally
irreducible in all bipartitions? Intuition suggests that a genuinely
entangled orthogonal basis (the basis vectors are entangled in every
bipartition) is a promising candidate because in any bipartition,
the states are not only locally indistinguishable but also none can
be correctly identified with a nonzero probability using LOCC \cite{HSSH}.
However, we find that the GHZ basis, which is genuinely entangled
and locally irreducible {[}Proposition \ref{2X2X2-GHZ}{]}, is locally
reducible in all bipartitions. 
\begin{prop}
\label{GHZ-cut-decomposable} The three-qubit GHZ basis given in proposition
\ref{2X2X2-GHZ} is locally reducible in all bipartitions. 
\end{prop}
The proof is simple. Note that, one can always perform a joint measurement
on any two qubits to distinguish the subspaces spanned by $\left\{ \left|00\right\rangle ,\left|11\right\rangle \right\} $
and $\left\{ \left|01\right\rangle ,\left|10\right\rangle \right\} $.
Thus in any bipartition, the whole set can be locally reduced to two
disjoint subsets, each of which is locally equivalent to the Bell
basis (the proof can be extended \emph{mutatis mutandis} for a $N$-qubit
GHZ basis with the identical conclusion).

Proposition \ref{GHZ-cut-decomposable} gives rise to an interesting
question: Can multiparty orthogonal product states be locally irreducible
in all bipartitions? If such sets exist, then they would clearly demonstrate\emph{
}quantum nonlocality stronger than what we presently understand\emph{.}

First we observe that such product states cannot be found in systems
where one of subsystems has dimension two: If the system contains
a qubit, then the set is locally distinguishable in the bipartition
${\bf qubit}\vert{\bf rest}$ because orthogonal product states on
$\mathbb{C}^{2}\otimes\mathbb{C}^{d}$, $d\geq2$ are known to be
locally distinguishable \cite{divin03}. So they can only exist, if
at all, on $\mathcal{H}=\bigotimes_{i=1}^{n}\mathcal{H}_{i}$, $n\geq3$,
where $\dim\mathcal{H}_{i}\geq3$ for every $i$. Thus the minimum
dimension corresponds to a three-qutrit system. 

We checked all the known examples (to the best of our knowledge) of
locally indistinguishable multiparty orthogonal product states, but
did not find any with the desired property. Some were ruled out by
the dimensionality constraint, and the rest turned out to be either
locally distinguishable \cite{divin03,nis06,halder,Xu-17,Wang-2017-Qinfoprocess,zhang14,zhang15,wang17}
or locally reducible \cite{Zhang-Oh-2017} in one or more bipartitions.

The main result of this paper lies in showing that multiparty orthogonal
product states that are locally irreducible in all bipartitions, exist.
We call such sets strongly nonlocal. 
\begin{defn}
\label{strong-nonlocality} Consider a composite quantum system $\mathcal{H}=\bigotimes_{i=1}^{n}\mathcal{H}_{i}$
with $n\geq3$ and $\dim\mathcal{H}_{i}\geq3$, $i=1,\dots,n$. A
set of orthogonal product states $\left|\psi_{i}\right\rangle =\left|\alpha_{i}\right\rangle _{1}\otimes\left|\beta_{i}\right\rangle _{2}\otimes\cdots\otimes\left|\gamma_{i}\right\rangle _{n}$
on $\mathcal{H}$ is strongly nonlocal if it is locally irreducible
in every bipartition. 
\end{defn}
We now give an example of an OPB on $\mathbb{C}^{3}\otimes\mathbb{C}^{3}\otimes\mathbb{C}^{3}$
and prove it strongly nonlocal. Note that, this construction achieves
the minimum dimension required (as discussed earlier). We will use
the notation $\left|1\right\rangle $, $\left|2\right\rangle $, $\left|3\right\rangle $
for the bases of Alice, Bob, and Charlie's Hilbert spaces. Consider
the following OPB on $\mathbb{C}^{3}\otimes\mathbb{C}^{3}\otimes\mathbb{C}^{3}$:
\begin{equation}
\begin{array}{ccccccc}
\left|1\right\rangle \left|2\right\rangle \left|1\pm2\right\rangle  &  &  & \left|2\right\rangle \left|1\pm2\right\rangle \left|1\right\rangle  &  &  & \left|1\pm2\right\rangle \left|1\right\rangle \left|2\right\rangle \\
\left|1\right\rangle \left|3\right\rangle \left|1\pm3\right\rangle  &  &  & \left|3\right\rangle \left|1\pm3\right\rangle \left|1\right\rangle  &  &  & \left|1\pm3\right\rangle \left|1\right\rangle \left|3\right\rangle \\
\left|2\right\rangle \left|3\right\rangle \left|1\pm2\right\rangle  &  &  & \left|3\right\rangle \left|1\pm2\right\rangle \left|2\right\rangle  &  &  & \left|1\pm2\right\rangle \left|2\right\rangle \left|3\right\rangle \\
\left|3\right\rangle \left|2\right\rangle \left|1\pm3\right\rangle  &  &  & \left|2\right\rangle \left|1\pm3\right\rangle \left|3\right\rangle  &  &  & \left|1\pm3\right\rangle \left|3\right\rangle \left|2\right\rangle \\
\left|1\right\rangle \left|1\right\rangle \left|1\right\rangle  &  &  & \left|2\right\rangle \left|2\right\rangle \left|2\right\rangle  &  &  & \left|3\right\rangle \left|3\right\rangle \left|3\right\rangle ,
\end{array}\label{eq:COPB3X3X3}
\end{equation}
where $\left|1\pm2\right\rangle $ stands for $\frac{1}{\sqrt{2}}\left(\left|1\right\rangle \pm\left|2\right\rangle \right)$
etc. Note that, the set (\ref{eq:COPB3X3X3}) is invariant under cyclic
permutation of the parties $A$, $B$, and $C$. We first show that
the states are locally irreducible. 
\begin{lem}
\label{3X3X3-locally-indecomposable} The set of states given by (\ref{eq:COPB3X3X3})
on $\mathbb{C}^{3}\otimes\mathbb{C}^{3}\otimes\mathbb{C}^{3}$ is
locally irreducible. 
\end{lem}
To prove the lemma, we first consider the following states 
\begin{equation}
\begin{array}{ccccccc}
\left|1\right\rangle \left|2\right\rangle \left|1\pm2\right\rangle  &  &  & \left|2\right\rangle \left|1\pm2\right\rangle \left|1\right\rangle  &  &  & \left|1\pm2\right\rangle \left|1\right\rangle \left|2\right\rangle \\
\left|1\right\rangle \left|3\right\rangle \left|1\pm3\right\rangle  &  &  & \left|3\right\rangle \left|1\pm3\right\rangle \left|1\right\rangle  &  &  & \left|1\pm3\right\rangle \left|1\right\rangle \left|3\right\rangle ,
\end{array}\label{small-subset-3X3X3}
\end{equation}
chosen from the whole set. For the above states it was shown \cite{halder}
that any $3\times3$ orthogonality-preserving POVM acting on any subsystem
must be proportional to the identity. Clearly, this must also hold
for the whole set (\ref{eq:COPB3X3X3}) of which the states (\ref{small-subset-3X3X3})
form a subset because all the states belong to the same state space
$\mathbb{C}^{3}\otimes\mathbb{C}^{3}\otimes\mathbb{C}^{3}$. Therefore,
none of the parties can begin a LOCC protocol by performing some local
NOPM. Hence, the proof (for completeness, we have included the details
in Appendix E). 
\begin{thm}
The orthogonal product basis (\ref{eq:COPB3X3X3}) is strongly nonlocal. 
\end{thm}
We need to show that the states (\ref{eq:COPB3X3X3}) form a locally
irreducible set in any bipartition. To begin with, consider the bipartition
$A\vert BC$ ($\mathcal{H}_{A}\otimes\mathcal{H}_{BC}$). In this
bipartition the states (\ref{eq:COPB3X3X3}) take the form 
\begin{equation}
\begin{array}{ccccccc}
\left|1\right\rangle \left|21\pm22\right\rangle  &  &  & \left|2\right\rangle \left|11\pm21\right\rangle  &  &  & \left|1\pm2\right\rangle \left|12\right\rangle \\
\left|1\right\rangle \left|31\pm33\right\rangle  &  &  & \left|3\right\rangle \left|11\pm31\right\rangle  &  &  & \left|1\pm3\right\rangle \left|13\right\rangle \\
\left|2\right\rangle \left|31\pm32\right\rangle  &  &  & \left|3\right\rangle \left|12\pm22\right\rangle  &  &  & \left|1\pm2\right\rangle \left|23\right\rangle \\
\left|3\right\rangle \left|21\pm23\right\rangle  &  &  & \left|2\right\rangle \left|13\pm33\right\rangle  &  &  & \left|1\pm3\right\rangle \left|32\right\rangle \\
\left|1\right\rangle \left|11\right\rangle  &  &  & \left|2\right\rangle \left|22\right\rangle  &  &  & \left|3\right\rangle \left|33\right\rangle .
\end{array}\label{3X9without mapping}
\end{equation}
Physically this means the subsystems $B$ and $C$ are treated together
as a nine-dimensional subsystem $BC$. For clarity, denote the elements
of the basis $\left\{ \left|ij\right\rangle \right\} _{i,j=1}^{3}$
on $\mathcal{H}_{BC}$ as: $\forall i=1,2,3$, $\left|1i\right\rangle \rightarrow\mbox{\ensuremath{\left|\mathbf{i}\right\rangle }},\left|2i\right\rangle \rightarrow\mbox{\ensuremath{\left|\mathbf{i+3}\right\rangle }}$,
and $\left|3i\right\rangle \rightarrow\mbox{\ensuremath{\left|\mathbf{i+6}\right\rangle }}$
and rewrite the states (\ref{3X9without mapping}) as: 
\begin{equation}
\begin{array}{ccccccc}
\left|1\right\rangle \left|{\bf 4}\pm{\bf 5}\right\rangle  &  &  & \left|2\right\rangle \left|{\bf 1}\pm{\bf 4}\right\rangle  &  &  & \left|1\pm2\right\rangle \left|{\bf 2}\right\rangle \\
\left|1\right\rangle \left|\mathbf{7}\pm\mathbf{9}\right\rangle  &  &  & \left|3\right\rangle \left|\mathbf{1}\pm\mathbf{7}\right\rangle  &  &  & \left|1\pm3\right\rangle \left|\mathbf{3}\right\rangle \\
\left|2\right\rangle \left|\mathbf{7}\pm\mathbf{8}\right\rangle  &  &  & \left|3\right\rangle \left|\mathbf{2}\pm\mathbf{5}\right\rangle  &  &  & \left|1\pm2\right\rangle \left|\mathbf{6}\right\rangle \\
\left|3\right\rangle \left|\mathbf{4}\pm\mathbf{6}\right\rangle  &  &  & \left|2\right\rangle \left|\mathbf{3}\pm\mathbf{9}\right\rangle  &  &  & \left|1\pm3\right\rangle \left|\mathbf{8}\right\rangle \\
\left|1\right\rangle \left|\mathbf{1}\right\rangle  &  &  & \left|2\right\rangle \left|\mathbf{5}\right\rangle  &  &  & \left|3\right\rangle \left|\mathbf{9}\right\rangle .
\end{array}\label{COPB3X9}
\end{equation}
We now show that any orthogonality-preserving local POVM performed
either on $A$ or $BC$ must be trivial. Therefore, neither Alice
($A$) nor Bob and Charlie together ($BC)$ can go first.

First, consider Alice. Recall that, Lemma \ref{3X3X3-locally-indecomposable}
holds because none of the parties can perform a local NOPM when all
parts are separated. Since in the bipartition $A\vert BC$ Alice's
subsystem is still separated from the rest, we conclude that Alice
cannot go first.

We now consider whether it is possible to initiate a local protocol
by performing some NOPM on $BC$. Let the POVM $\left\{ \Pi_{\alpha}\right\} $
describe a general orthogonality-preserving measurement on $BC$.
Each POVM element $\Pi_{\alpha}$ can be written as a $9\times9$
matrix in the $\left\{ \left|\mathbf{1}\right\rangle ,\dots,\left|\mathbf{9}\right\rangle \right\} $
basis of $\mathcal{H}_{BC}$: 
\begin{eqnarray}
\Pi_{\alpha} & = & \begin{pmatrix}a_{11} & a_{12} & a_{13} & a_{14} & a_{15} & a_{16} & a_{17} & a_{18} & a_{19}\\
a_{21} & a_{22} & a_{23} & a_{24} & a_{25} & a_{26} & a_{27} & a_{28} & a_{29}\\
a_{31} & a_{32} & a_{33} & a_{34} & a_{35} & a_{36} & a_{37} & a_{38} & a_{39}\\
a_{41} & a_{42} & a_{43} & a_{44} & a_{45} & a_{46} & a_{47} & a_{48} & a_{49}\\
a_{51} & a_{52} & a_{53} & a_{54} & a_{55} & a_{56} & a_{57} & a_{58} & a_{59}\\
a_{61} & a_{62} & a_{63} & a_{64} & a_{65} & a_{66} & a_{67} & a_{68} & a_{69}\\
a_{71} & a_{72} & a_{73} & a_{74} & a_{75} & a_{76} & a_{77} & a_{78} & a_{79}\\
a_{81} & a_{82} & a_{83} & a_{84} & a_{85} & a_{86} & a_{87} & a_{88} & a_{89}\\
a_{91} & a_{92} & a_{93} & a_{94} & a_{95} & a_{96} & a_{97} & a_{98} & a_{99}
\end{pmatrix}.\label{Pi-alpha-matrix}
\end{eqnarray}
The measurement must leave the postmeasurement states mutually orthogonal.
By choosing suitable pairs of vectors $\left\{ \left|\psi_{i}\right\rangle ,\left|\psi_{j}\right\rangle \right\} $,
$i\neq j$, we find that all the off-diagonal matrix elements $a_{ij}$,
$i\neq j$ must be zero if the orthogonality-preserving conditions
$\left\langle \psi_{i}\left|\mathbb{I}\otimes\Pi_{\alpha}\right|\psi_{j}\right\rangle =0$
are to be satisfied. Table I in Appendix F shows the complete analysis.
Similarly, we find that the diagonal elements are all equal. For example,
by setting the inner product $\langle1|\langle\mathbf{4}+\mathbf{5}|\mathbb{I}\otimes\Pi_{\alpha}|1\rangle|\mathbf{4}-\mathbf{5}\rangle=0$,
we get $a_{44}$ = $a_{55}$. Table II (Appendix F) summarizes this
analysis. As the diagonal elements of $\Pi_{\alpha}$ are all equal
and the off-diagonal elements are all zero, $\Pi_{\alpha}$ must be
proportional to the identity. The argument applies to all measurement
outcomes, and thus all POVM elements $\left\{ \Pi_{\alpha}\right\} $
must be proportional to the identity. This means the POVM must be
trivial, and therefore, $BC$ cannot go first. Thus the states (\ref{COPB3X9})
form a locally irreducible set in the bipartition $A\vert BC$. Now
from the symmetry of the states (\ref{eq:COPB3X3X3}) {[}invariant
under cyclic permutation of the parties{]}, it follows that the states
(\ref{eq:COPB3X3X3}) are also locally irreducible in the bipartitions
$C\vert AB$, and $B\vert CA$. This completes the proof of the theorem.

In Appendix G we have given an example of a strongly nonlocal OPB
on $\mathbb{C}^{4}\otimes\mathbb{C}^{4}\otimes\mathbb{C}^{4}$ along
with the complete proof.

We now discuss the question of local discrimination of strongly nonlocal
product states using entanglement as a resource. Note that, in our
examples, the three-party separable measurements cannot be locally
implemented even if any two share unlimited entanglement. So exact
local implementation would require a resource state which must be
entangled in all bipartitions (this also holds for product states
that are locally indistinguishable in all bipartitions \cite{Zhang-Oh-2017}
but not strongly nonlocal). As to how much entanglement must one consume,
we do not have any clear answer. A teleportation protocol can perfectly
distinguish the states (\ref{eq:COPB3X3X3}) using $\mathbb{C}^{3}\otimes\mathbb{C}^{3}$
maximally entangled states shared between any two pairs, but whether
one can do just as well using cheaper resources (see Ref. \cite{Cohen-2008})
is an intriguing question.

The results in this paper also leave open other interesting questions.
One may consider generalizing our constructions on $\bigotimes_{i=1}^{n}\mathbb{\mathbb{C}}^{d}$
for $n\geq4$, and $d\geq3$. Another problem worth considering is
whether incomplete orthogonal product bases can be strongly nonlocal,
e.g., can we have a strongly nonlocal UPB? Finally, one may ask, whether
one can find entangled bases that are locally irreducible in all bipartitions.
In view of Proposition \ref{GHZ-cut-decomposable}, it seems that
to satisfy ``local irreducibility in all bipartitions,'' the structure
of the states is likely to be more important than their entanglement.
Here, even examples in the simplest case of $\mathbb{C}^{2}\otimes\mathbb{C}^{2}\otimes\mathbb{C}^{2}$
can help us to understand this property better. 
\begin{acknowledgments}
M. B. acknowledges support through an INSPIRE-faculty position at
S. N. Bose National Center for Basic Sciences by the Department of
Science and Technology, Government of India. S. B. is supported in
part by SERB (Science and Engineering Research Board), DST, Govt.
of India through Project No. EMR/2015/002373. \end{acknowledgments}

\onecolumngrid

\newpage

\section*{APPENDIX\protect \\
}

\textbf{A.} Here we give the local protocol which shows that the set
of states {[}given by (\ref{locally-reducible (b)}){]} 
\[
\begin{array}{ccccccc}
\left|00\right\rangle \pm\left|11\right\rangle  &  & \left|02\right\rangle  &  & \left|20\right\rangle  &  & \left|22\right\rangle \\
\left|01\right\rangle \pm\left|10\right\rangle  &  & \left|12\right\rangle  &  & \left|21\right\rangle 
\end{array}
\]
is locally reducible, with only one subset (the Bell basis) being
locally indistinguishable. The protocol goes like this: Both Alice
and Bob perform the measurement that distinguishes the subspaces spanned
by $\left\{ \left|0\right\rangle ,\left|1\right\rangle \right\} $
and $\left\{ \left|2\right\rangle \right\} $. The following table
summarizes the outcomes and inferences. 

\begin{equation}
\begin{array}{|c|c|c|}
\hline \;\;\;{\rm {\bf Alice}}\rightarrow & {\rm subspace} & {\rm subspace}\\
{\rm {\bf Bob}} & \left\{ \left|0\right\rangle ,\left|1\right\rangle \right\}  & \left\{ \left|2\right\rangle \right\} \\
\downarrow &  & \\
\hline {\rm subspace} & {\rm Bell\; Basis} & \left|20\right\rangle /\left|21\right\rangle \\
\left\{ \left|0\right\rangle ,\left|1\right\rangle \right\}  & {\rm locally} & {\rm locally}\\
 & {\rm indistinguishable} & {\rm distinguishable}\\
\hline {\rm subspace} & \left|02\right\rangle /\left|12\right\rangle  & \left|22\right\rangle \\
\left\{ \left|2\right\rangle \right\}  & {\rm locally} & {\rm locally}\\
 & {\rm distinguishable} & {\rm distinguishable}
\\\hline \end{array}\label{3X9=00003Dwihout mapping-1}
\end{equation}
\textbf{}\\
\textbf{B.} Proof of irreducibility of the Bell basis (Proposition
2): Let $\pi_{\alpha}=M_{\alpha}^{\dagger}M_{\alpha}$, where $M_{\alpha}$
is the Krauss operator. Since the measurement is orthogonality-preserving,
for every $\alpha$, the following states must be pairwise orthogonal
to each other.

\begin{spacing}{1.2}
\begin{equation}
\begin{array}{c}
(\mathbb{I}\otimes M_{\alpha})|\psi_{1}\rangle=(\mathbb{I}\otimes M_{\alpha})(|00\rangle+|11\rangle)\\
(\mathbb{I}\otimes M_{\alpha})|\psi_{2}\rangle=(\mathbb{I}\otimes M_{\alpha})(|00\rangle-|11\rangle)\\
(\mathbb{I}\otimes M_{\alpha})|\psi_{3}\rangle=(\mathbb{I}\otimes M_{\alpha})(|01\rangle+|10\rangle)\\
(\mathbb{I}\otimes M_{\alpha})|\psi_{4}\rangle=(\mathbb{I}\otimes M_{\alpha})(|01\rangle-|10\rangle)
\end{array}
\end{equation}
Setting the inner products to zero, we solve for the matrix elements:

\begin{equation}
\begin{array}{|c|c|c|}
\hline \langle\psi_{1}|\mathbb{I}\otimes\pi_{\alpha}|\psi_{2}\rangle=0 & a_{00}-a_{11}=0 & a_{00}=a_{11}\\
\hline \langle\psi_{1}|\mathbb{I}\otimes\pi_{\alpha}|\psi_{3}\rangle=0 & a_{01}+a_{10}=0 & a_{01}=a_{10}=0\\
\langle\psi_{1}|\mathbb{I}\otimes\pi_{\alpha}|\psi_{4}\rangle=0 & a_{01}-a_{10}=0 & 
\\\hline \end{array}
\end{equation}
As we can see $\pi_{\alpha}$ must be proportional to the identity.
\\
\textbf{}\\
\textbf{C.} Proof of irreducibility of the $GHZ$ basis: The states
in the $GHZ$ basis are given by:

\begin{equation}
\begin{array}{c}
|G_{1}\rangle=|0\rangle|0\rangle|0\rangle+|1\rangle|1\rangle|1\rangle,\\
|G_{2}\rangle=|0\rangle|0\rangle|1\rangle+|1\rangle|1\rangle|0\rangle,\\
|G_{3}\rangle=|0\rangle|1\rangle|0\rangle+|1\rangle|0\rangle|1\rangle,\\
|G_{4}\rangle=|0\rangle|1\rangle|1\rangle+|1\rangle|0\rangle|0\rangle,\\
|G_{5}\rangle=|0\rangle|0\rangle|0\rangle-|1\rangle|1\rangle|1\rangle,\\
|G_{6}\rangle=|0\rangle|0\rangle|1\rangle-|1\rangle|1\rangle|0\rangle,\\
|G_{7}\rangle=|0\rangle|1\rangle|0\rangle-|1\rangle|0\rangle|1\rangle,\\
|G_{8}\rangle=|0\rangle|1\rangle|1\rangle-|1\rangle|0\rangle|0\rangle.
\end{array}
\end{equation}
The proof works exactly the same way as the previous one. Let's first
consider Charlie. A measurement by Charlie can be defined by a set
of POVM elements $\{\pi_{i}\}$, $\sum_{i}\pi_{i}=\mathbb{I}$. In
the $\{|0\rangle,|1\rangle\}$ basis the matrix form of $\pi_{i}$
= $M_{i}^{\dagger}M_{i}$ is given by: 
\begin{equation}
\begin{pmatrix}a_{00} & a_{01}\\
a_{10} & a_{11}
\end{pmatrix}
\end{equation}
For the outcome $i$, the post-measurement states are,

\begin{equation}
\begin{array}{c}
(\mathbb{I}\otimes\mathbb{I}\otimes M_{i})|G_{1}\rangle=(\mathbb{I}\otimes\mathbb{I}\otimes M_{i})(|0\rangle|0\rangle|0\rangle+|1\rangle|1\rangle|1\rangle),\\
(\mathbb{I}\otimes\mathbb{I}\otimes M_{i})|G_{2}\rangle=(\mathbb{I}\otimes\mathbb{I}\otimes M_{i})(|0\rangle|0\rangle|1\rangle+|1\rangle|1\rangle|0\rangle),\\
(\mathbb{I}\otimes\mathbb{I}\otimes M_{i})|G_{3}\rangle=(\mathbb{I}\otimes\mathbb{I}\otimes M_{i})(|0\rangle|1\rangle|0\rangle+|1\rangle|0\rangle|1\rangle),\\
(\mathbb{I}\otimes\mathbb{I}\otimes M_{i})|G_{4}\rangle=(\mathbb{I}\otimes\mathbb{I}\otimes M_{i})(|0\rangle|1\rangle|1\rangle+|1\rangle|0\rangle|0\rangle),\\
(\mathbb{I}\otimes\mathbb{I}\otimes M_{i})|G_{5}\rangle=(\mathbb{I}\otimes\mathbb{I}\otimes M_{i})(|0\rangle|0\rangle|0\rangle-|1\rangle|1\rangle|1\rangle),\\
(\mathbb{I}\otimes\mathbb{I}\otimes M_{i})|G_{6}\rangle=(\mathbb{I}\otimes\mathbb{I}\otimes M_{i})(|0\rangle|0\rangle|1\rangle-|1\rangle|1\rangle|0\rangle),\\
(\mathbb{I}\otimes\mathbb{I}\otimes M_{i})|G_{7}\rangle=(\mathbb{I}\otimes\mathbb{I}\otimes M_{i})(|0\rangle|1\rangle|0\rangle-|1\rangle|0\rangle|1\rangle),\\
(\mathbb{I}\otimes\mathbb{I}\otimes M_{i})|G_{8}\rangle=(\mathbb{I}\otimes\mathbb{I}\otimes M_{i})(|0\rangle|1\rangle|1\rangle-|1\rangle|0\rangle|0\rangle).
\end{array}
\end{equation}
As the measurement is orthogonality-preserving, the above states must
be mutually orthogonal. Setting the inner products to zero we can
easily solve for the matrix elements:

\begin{equation}
\begin{array}{|c|c|c|}
\hline \langle G_{1}|\mathbb{I}\otimes\mathbb{I}\otimes\pi_{i}|G_{5}\rangle=0 & a_{00}-a_{11}=0 & a_{00}=a_{11}\\
\hline \langle G_{1}|\mathbb{I}\otimes\mathbb{I}\otimes\pi_{i}|G_{2}\rangle=0 & a_{01}+a_{10}=0 & a_{01}=a_{10}=0\\
\langle G_{1}|\mathbb{I}\otimes\mathbb{I}\otimes\pi_{i}|G_{6}\rangle=0 & a_{01}-a_{10}=0 & 
\\\hline \end{array}
\end{equation}
Thus $\pi_{i}$ is proportional to a $2\times2$ identity matrix.
Since the argument applies to all possible outcomes, Charlie cannot
go first and from the symmetry of the states, neither can Alice or
Bob. \\
\\
\textbf{D.} \label{Incomplete GHZ} Consider the following $GHZ$
states (unnormalized) on $\mathbb{C}^{2}\otimes\mathbb{C}^{2}\otimes\mathbb{C}^{2}$
\begin{equation}
\begin{array}{c}
\left|0\right\rangle _{A}\left|0\right\rangle _{B}\left|0\right\rangle _{C}\pm\left|1\right\rangle _{A}\left|1\right\rangle _{B}\left|1\right\rangle _{C}\\
\left|0\right\rangle _{A}\left|1\right\rangle _{B}\left|1\right\rangle _{C}\pm\left|1\right\rangle _{A}\left|0\right\rangle _{B}\left|0\right\rangle _{C}
\end{array}\label{incomplete-GHZ}
\end{equation}
The above states cannot be distinguished across $A\vert BC$ (locally
equivalent to the Bell basis) but are distinguishable across the other
two bipartitions $B\vert CA$ and $C\vert AB$. \\
\\
\textbf{E.} Proof of Lemma 1: The twelve product states are given
by:

\begin{equation}
\begin{array}{cc}
|\psi_{1}\rangle=|1\rangle|2\rangle|1+2\rangle, & |\psi_{2}\rangle=|1\rangle|2\rangle|1-2\rangle,\\[0.5ex]
|\psi_{3}\rangle=|1\rangle|3\rangle|1+3\rangle, & |\psi_{4}\rangle=|1\rangle|3\rangle|1-3\rangle,\\[0.5ex]
|\psi_{5}\rangle=|2\rangle|1+2\rangle|1\rangle, & |\psi_{6}\rangle=|2\rangle|1-2\rangle|1\rangle,\\[0.5ex]
|\psi_{7}\rangle=|3\rangle|1+3\rangle|1\rangle, & |\psi_{8}\rangle=|3\rangle|1-3\rangle|1\rangle,\\[0.5ex]
|\psi_{9}\rangle=|1+2\rangle|1\rangle|2\rangle, & |\psi_{10}\rangle=|1-2\rangle|1\rangle|2\rangle,\\[0.5ex]
|\psi_{11}\rangle=|1+3\rangle|1\rangle|3\rangle, & |\psi_{12}\rangle=|1-3\rangle|1\rangle|3\rangle.
\end{array}
\end{equation}
Suppose that Alice goes first. Her measurement is defined by a set
of POVM elements $\{\pi_{l}\}$, $\sum_{l}\pi_{l}$ = $\mathbb{I}_{3\times3}$.
In matrix form, $\pi_{l}$ = $M_{l}^{\dagger}M_{l}$ can be written
as (in \{$|1\rangle$, $|2\rangle$, $|3\rangle$\} basis)

\begin{equation}
\pi_{l}=M_{l}^{\dagger}M_{l}=\begin{pmatrix}e_{11} & e_{12} & e_{13}\\
e_{21} & e_{22} & e_{23}\\
e_{31} & e_{32} & e_{33}
\end{pmatrix}.
\end{equation}
As the measurement is orthogonality preserving then after any given
outcome, say $l,$ the post measurement states $(M_{l}\otimes\mathbb{I}\otimes\mathbb{I})|\psi_{k}\rangle$,
$k=1,\dots,12$ remain pairwise orthogonal to each other. Setting
the inner products of the post measurement states equal to zero, we
solve for the matrix elements:

\begin{equation}
\begin{array}{|c|c|}
\hline \langle\psi_{5}|\pi_{l}\otimes\mathbb{I}\otimes\mathbb{I}|\psi_{7}\rangle=0 & e_{23}=0\\
\langle\psi_{7}|\pi_{l}\otimes\mathbb{I}\otimes\mathbb{I}|\psi_{5}\rangle=0 & e_{32}=0\\
\hline \langle\psi_{1}|\pi_{l}\otimes\mathbb{I}\otimes\mathbb{I}|\psi_{5}\rangle=0 & e_{12}=0\\
\langle\psi_{5}|\pi_{l}\otimes\mathbb{I}\otimes\mathbb{I}|\psi_{1}\rangle=0 & e_{21}=0\\
\hline \langle\psi_{3}|\pi_{l}\otimes\mathbb{I}\otimes\mathbb{I}|\psi_{7}\rangle=0 & e_{13}=0\\
\langle\psi_{7}|\pi_{l}\otimes\mathbb{I}\otimes\mathbb{I}|\psi_{3}\rangle=0 & e_{31}=0\\
\hline \langle\psi_{9}|\pi_{l}\otimes\mathbb{I}\otimes\mathbb{I}|\psi_{10}\rangle=0 & e_{11}=e_{22}\\
\langle\psi_{11}|\pi_{l}\otimes\mathbb{I}\otimes\mathbb{I}|\psi_{12}\rangle=0 & e_{11}=e_{33}
\\\hline \end{array}
\end{equation}
Thus we see that $\pi_{l}$ is proportional to the $3\times3$ identity
matrix. Since the argument holds for all outcomes, Alice cannot go
first and from the symmetry neither can Bob or Charlie. \\
\textbf{}\\
\textbf{F. Tables I and II}
\end{spacing}

\begin{table}[ht]
\protect\caption{Off-diagonal elements}

\centering %
\begin{tabular}{|c|c|c||c|c|c|}
\hline 
Sl. No.  & States  & Elements  & Sl. No.  & States  & Elements \tabularnewline
\hline 
\hline 
(1) & $|1+2\rangle|{\bf 2}\rangle$, $|1\rangle|{\bf 1}\rangle$  & $a_{21}$ = $a_{12}$ = 0  & (2)  & $|1\rangle|{\bf 1}\rangle$, $|1+3\rangle|{\bf 3}\rangle$  & $a_{31}$ = $a_{13}$ = 0 \tabularnewline
\hline 
(3)  & $|1\rangle|{\bf 1}\rangle$, $|1+2\rangle|{\bf 6}\rangle$  & $a_{61}$ = $a_{16}$ = 0  & (4)  & $|1\rangle|{\bf 1}\rangle$, $|1+3\rangle|{\bf 8}\rangle$  & $a_{81}$ = $a_{18}$ = 0 \tabularnewline
\hline 
(5)  & $|2\rangle|{\bf 5}\rangle$, $|1+2\rangle|{\bf 2}\rangle$  & $a_{52}$ = $a_{25}$ = 0  & (6)  & $|2\rangle|{\bf 5}\rangle$, $|1+2\rangle|{\bf 6}\rangle$  & $a_{65}$ = $a_{56}$ = 0 \tabularnewline
\hline 
(7)  & $|3\rangle|{\bf 9}\rangle$, $|1+3\rangle|{\bf 3}\rangle$  & $a_{93}$ = $a_{39}$ = 0  & (8)  & $|3\rangle|{\bf 9}\rangle$, $|1+3\rangle|{\bf 8}\rangle$  & $a_{98}$ = $a_{89}$ = 0 \tabularnewline
\hline 
(9)  & $|1\rangle|{\bf 4}+{\bf 5}\rangle$, $|1+2\rangle|{\bf 2}\rangle$  & $a_{42}$ = $a_{24}$ = 0  & (10)  & $|1\rangle|{\bf 4}+{\bf 5}\rangle$, $|1+2\rangle|{\bf 6}\rangle$  & $a_{46}$ = $a_{64}$ = 0 \tabularnewline
\hline 
(11)  & $|1+2\rangle|{\bf 2}\rangle$, $|1+3\rangle|{\bf 3}\rangle$  & $a_{32}$ = $a_{23}$ = 0  & (12)  & $|1+2\rangle|{\bf 2}\rangle$, $|1+2\rangle|{\bf 6}\rangle$  & $a_{62}$ = $a_{26}$ = 0 \tabularnewline
\hline 
(13)  & $|1+2\rangle|{\bf 2}\rangle$, $|1+3\rangle|{\bf 8}\rangle$  & $a_{82}$ = $a_{28}$ = 0  & (14)  & $|1+2\rangle|{\bf 2}\rangle$, $|2\rangle|{\bf 7}+{\bf 8}\rangle$  & $a_{72}$ = $a_{27}$ = 0 \tabularnewline
\hline 
(15)  & $|1\rangle|{\bf 7}+{\bf 9}\rangle$, $|1+3\rangle|{\bf 3}\rangle$  & $a_{73}$ = $a_{37}$ = 0  & (16)  & $|1\rangle|{\bf 7}+{\bf 9}\rangle$, $|1+3\rangle|{\bf 8}\rangle$  & $a_{87}$ = $a_{78}$ = 0 \tabularnewline
\hline 
(17)  & $|1+2\rangle|{\bf 2}\rangle$, $|1\rangle|{\bf 7}+{\bf 9}\rangle$  & $a_{92}$ = $a_{29}$ = 0  & (18)  & $|1+3\rangle|{\bf 3}\rangle$, $|3\rangle|{\bf 2}+{\bf 5}\rangle$  & $a_{53}$ = $a_{35}$ = 0 \tabularnewline
\hline 
(19)  & $|1+3\rangle|{\bf 3}\rangle$, $|1+2\rangle|{\bf 6}\rangle$  & $a_{63}$ = $a_{36}$ = 0  & (20)  & $|1+3\rangle|{\bf 3}\rangle$, $|1+3\rangle|{\bf 8}\rangle$  & $a_{83}$ = $a_{38}$ = 0 \tabularnewline
\hline 
(21)  & $|1+3\rangle|{\bf 3}\rangle$, $|3\rangle|{\bf 4}+{\bf 6}\rangle$  & $a_{43}$ = $a_{34}$ = 0  & (22)  & $|1+2\rangle|{\bf 6}\rangle$, $|1+3\rangle|{\bf 8}\rangle$  & $a_{86}$ = $a_{68}$ = 0 \tabularnewline
\hline 
(23)  & $|1+2\rangle|{\bf 6}\rangle$, $|2\rangle|{\bf 7}+{\bf 8}\rangle$  & $a_{76}$ = $a_{67}$ = 0  & (24)  & $|1+2\rangle|{\bf 6}\rangle$, $|1\rangle|{\bf 7}+{\bf 9}\rangle$  & $a_{96}$ = $a_{69}$ = 0 \tabularnewline
\hline 
(25)  & $|1+3\rangle|{\bf 8}\rangle$, $|3\rangle|{\bf 2}+{\bf 5}\rangle$  & $a_{85}$ = $a_{58}$ = 0  & (26)  & $|1+3\rangle|{\bf 8}\rangle$, $|1\rangle|{\bf 4}+{\bf 5}\rangle$  & $a_{84}$ = $a_{48}$ = 0 \tabularnewline
\hline 
(27)  & $|3\rangle|{\bf 2}+{\bf 5}\rangle$, $|3\rangle|{\bf 4}+{\bf 6}\rangle$  & $a_{54}$ = $a_{45}$ = 0  & (28)  & $|2\rangle|{\bf 5}\rangle$, $|2\rangle|{\bf 1}+{\bf 4}\rangle$  & $a_{51}$ = $a_{15}$ = 0 \tabularnewline
\hline 
(29)  & $|2\rangle|{\bf 5}\rangle$, $|2\rangle|{\bf 3}+{\bf 9}\rangle$  & $a_{95}$ = $a_{59}$ = 0  & (30)  & $|3\rangle|{\bf 1}+{\bf 7}\rangle$, $|3\rangle|{\bf 2}+{\bf 5}\rangle$  & $a_{75}$ = $a_{57}$ = 0 \tabularnewline
\hline 
(31)  & $|2\rangle|{\bf 3}+{\bf 9}\rangle$, $|2\rangle|{\bf 7}+{\bf 8}\rangle$  & $a_{97}$ = $a_{79}$ = 0  & (32)  & $|3\rangle|{\bf 1}+{\bf 7}\rangle$, $|3\rangle|{\bf 9}\rangle$  & $a_{91}$ = $a_{19}$ = 0 \tabularnewline
\hline 
(33)  & $|3\rangle|{\bf 4}+{\bf 6}\rangle$, $|3\rangle|{\bf 9}\rangle$  & $a_{94}$ = $a_{49}$ = 0  & (34)  & $|1\rangle|{\bf 4}+{\bf 5}\rangle$, $|1\rangle|{\bf 7}+{\bf 9}\rangle$  & $a_{74}$ = $a_{47}$ = 0 \tabularnewline
\hline 
(35)  & $|3\rangle|{\bf 1}+{\bf 7}\rangle$, $|3\rangle|{\bf 4}+{\bf 6}\rangle$  & $a_{41}$ = $a_{14}$ = 0  & (36)  & $|1\rangle|{\bf 7}+\mathbf{9}\rangle$, $|1\rangle|{\bf 1}\rangle$  & $a_{71}$ = $a_{17}$ = 0 \tabularnewline
\hline 
\end{tabular}
\end{table}

Note: The analysis was done according to the serial numbers, and that's
how the table should be read/followed. In some cases, the inner product
conditions give us the values of $a_{ij}$s, $i\neq j$ right away.
For example, consider entry (1): Here, for the pair of states $\left\{ \left|1+2\right\rangle \left|\mathbf{2}\right\rangle ,\left|1\right\rangle \left|\mathbf{1}\right\rangle \right\} $,
the conditions $\left\langle 1+2\vert1\right\rangle \left\langle \mathbf{2}\left|\Pi_{\alpha}\right|\mathbf{1}\right\rangle =0$
and $\left\langle 1\vert1+2\right\rangle \left\langle \mathbf{1}\left|\Pi_{\alpha}\right|\mathbf{2}\right\rangle =0$
yield $a_{21}=a_{12}=0$. In some other cases, e.g. entry (9) the
inner-product conditions give the equations: (a) $a_{42}+a_{52}=0$
and (b) $a_{24}+a_{25}=0$ which can't be solved directly. But in
entry (5) we have already obtained the values for $a_{25}$ and $a_{52}$,
both of which are zero. Therefore, $a_{42}=a_{24}=0$. \\

\begin{table}[ht]
\protect\caption{Diagonal elements }

\centering %
\begin{tabular}{|c|c||c|c|}
\hline 
states  & elements  & states  & elements \tabularnewline
\hline 
\hline 
$|1\rangle|{\bf 4}+{\bf 5}\rangle$, $|1\rangle|{\bf 4}-{\bf 5}\rangle$  & $a_{44}$ = $a_{55}$  & $|2\rangle|{\bf 1}+{\bf 4}\rangle$, $|2\rangle|{\bf 1}-{\bf 4}\rangle$  & $a_{11}$ = $a_{44}$ \tabularnewline
\hline 
$|1\rangle|{\bf 7}+{\bf 9}\rangle$, $|1\rangle|{\bf 7}-{\bf 9}\rangle$  & $a_{77}$ = $a_{99}$  & $|3\rangle|{\bf 1}+{\bf 7}\rangle$, $|3\rangle|{\bf 1}-{\bf 7}\rangle$  & $a_{11}$ = $a_{77}$ \tabularnewline
\hline 
$|3\rangle|{\bf 2}+{\bf 5}\rangle$, $|3\rangle|{\bf 2}-{\bf 5}\rangle$  & $a_{22}$ = $a_{55}$  & $|2\rangle|{\bf 3}+{\bf 9}\rangle$, $|2\rangle|{\bf 3}-{\bf 9}\rangle$  & $a_{33}$ = $a_{99}$ \tabularnewline
\hline 
$|2\rangle|{\bf 7}+{\bf 8}\rangle$, $|2\rangle|{\bf 7}-{\bf 8}\rangle$  & $a_{77}$ = $a_{88}$  & $|3\rangle|{\bf 4}+{\bf 6}\rangle$, $|3\rangle|{\bf 4}-{\bf 6}\rangle$  & $a_{44}$ = $a_{66}$ \tabularnewline
\hline 
\end{tabular}
\end{table}
\textbf{}\\
\textbf{G. Strongly nonlocal OPB on $\mathbb{C}^{4}\otimes\mathbb{C}^{4}\otimes\mathbb{C}^{4}$ }

First, we obtain a small set of states similar to (\ref{small-subset-3X3X3});
there are eighteen such states:

\begin{equation}
S_{1}=\begin{array}{|ccc|}
\hline \;|1\rangle|2\rangle|1\pm2\rangle\;\; & |2\rangle|1\pm2\rangle|1\rangle\;\; & |1\pm2\rangle|1\rangle|2\rangle\;\\
\;|1\rangle|3\rangle|1\pm3\rangle\;\; & |3\rangle|1\pm3\rangle|1\rangle\;\; & |1\pm3\rangle|1\rangle|3\rangle\;\\
\;|1\rangle|4\rangle|1\pm4\rangle\;\; & |4\rangle|1\pm4\rangle|1\rangle\;\; & |1\pm4\rangle|1\rangle|4\rangle\;
\\\hline \end{array}\label{4 X 4 X 4 --block1}
\end{equation}
As the above states are locally distinguishable in bipartitions, we
add the twisted states given below:

\begin{equation}
S_{2}=\begin{array}{|ccc|}
\hline \;|2\rangle|3\rangle|1\pm2\rangle\;\; & |3\rangle|1\pm2\rangle|2\rangle\;\; & |1\pm2\rangle|2\rangle|3\rangle\;\\
\;|2\rangle|4\rangle|1\pm2\rangle\;\; & |4\rangle|1\pm2\rangle|2\rangle\;\; & |1\pm2\rangle|2\rangle|4\rangle\;\\
\;|3\rangle|4\rangle|1\pm3\rangle\;\; & |4\rangle|1\pm3\rangle|3\rangle\;\; & |1\pm3\rangle|3\rangle|4\rangle\;\\
\hline \;|4\rangle|3\rangle|1\pm4\rangle\;\; & |3\rangle|1\pm4\rangle|4\rangle\;\; & |1\pm4\rangle|4\rangle|3\rangle\;\\
\;|4\rangle|2\rangle|1\pm4\rangle\;\; & |2\rangle|1\pm4\rangle|4\rangle\;\; & |1\pm4\rangle|4\rangle|2\rangle\;\\
\;|3\rangle|2\rangle|1\pm3\rangle\;\; & |2\rangle|1\pm3\rangle|3\rangle\;\; & |1\pm3\rangle|3\rangle|2\rangle\;
\\\hline \end{array}\label{4 X 4 X4 block 2}
\end{equation}
Note that, in both blocks the second and third columns are obtained
by cyclic permutation of the first column {[}similar property holds
for (\ref{eq:COPB3X3X3}){]}. We can now complete the above set by
adding the \textquotedbl{}simple product\textquotedbl{} states--in
this case there are ten of them:

\begin{equation}
S_{3}=\left\{ \begin{array}{ccccc}
|2\rangle|3\rangle|4\rangle, & |3\rangle|4\rangle|2\rangle, & |4\rangle|2\rangle|3\rangle, & |2\rangle|4\rangle|3\rangle, & |4\rangle|3\rangle|2\rangle,\\
|3\rangle|2\rangle|4\rangle, & |1\rangle|1\rangle|1\rangle, & |2\rangle|2\rangle|2\rangle, & |3\rangle|3\rangle|3\rangle, & |4\rangle|4\rangle|4\rangle.
\end{array}\right\} \label{4 X 4 X 4 block 3}
\end{equation}
The union of the above three sets $S=S_{1}\cup S_{2}\cup S_{3}$ form
the desired OPB on $\mathbb{C}^{4}\otimes\mathbb{C}^{4}\otimes\mathbb{C}^{4}$.
In what follows, we first show that $S$ is locally irreducible and
then we will prove that $S$ is strongly nonlocal. 

First we show that $S$ is locally irreducible. To show this, we first
prove that that $S_{1}$ is locally irreducible. We write the states
(\ref{4 X 4 X 4 --block1} ) as: 

\begin{spacing}{1.2}
\begin{equation}
\begin{array}{cc}
|\psi_{1}\rangle=|1\rangle|2\rangle|1+2\rangle, & |\psi_{2}\rangle=|1\rangle|2\rangle|1-2\rangle,\\[0.5ex]
|\psi_{3}\rangle=|1\rangle|3\rangle|1+3\rangle, & |\psi_{4}\rangle=|1\rangle|3\rangle|1-3\rangle,\\[0.5ex]
|\psi_{5}\rangle=|1\rangle|4\rangle|1+4\rangle, & |\psi_{6}\rangle=|1\rangle|4\rangle|1-4\rangle,\\[0.5ex]
|\psi_{7}\rangle=|2\rangle|1+2\rangle|1\rangle, & |\psi_{8}\rangle=|2\rangle|1-2\rangle|1\rangle,\\[0.5ex]
|\psi_{9}\rangle=|3\rangle|1+3\rangle|1\rangle, & |\psi_{10}\rangle=|3\rangle|1-3\rangle|1\rangle,\\[0.5ex]
|\psi_{11}\rangle=|4\rangle|1+4\rangle|1\rangle, & |\psi_{12}\rangle=|4\rangle|1-4\rangle|1\rangle,\\[0.5ex]
|\psi_{13}\rangle=|1+2\rangle|1\rangle|2\rangle, & |\psi_{14}\rangle=|1-2\rangle|1\rangle|2\rangle,\\[0.5ex]
|\psi_{15}\rangle=|1+3\rangle|1\rangle|3\rangle, & |\psi_{16}\rangle=|1-3\rangle|1\rangle|3\rangle.\\[0.5ex]
|\psi_{17}\rangle=|1+4\rangle|1\rangle|4\rangle, & |\psi_{18}\rangle=|1-4\rangle|1\rangle|4\rangle.\\[0.5ex]
\end{array}
\end{equation}
Suppose Alice goes first. Alice's measurement is defined by a set
of POVM elements $\{\pi_{l}\}$, $\sum_{l}\pi_{l}$ = $\mathbb{I}_{4\times4}$.
Each element $\pi_{l}$ = $M_{l}^{\dagger}M_{l}$ is given by a $4\times4$
matrix written in \{$|1\rangle$, $|2\rangle$, $|3\rangle$, $|4\rangle$\}
basis: 

\begin{equation}
\pi_{l}=M_{l}^{\dagger}M_{l}=\begin{pmatrix}e_{11} & e_{12} & e_{13} & e_{14}\\
e_{21} & e_{22} & e_{23} & e_{24}\\
e_{31} & e_{32} & e_{33} & e_{34}\\
e_{41} & e_{42} & e_{43} & e_{44}
\end{pmatrix}.
\end{equation}
We assume that this measurement is orthogonality-preserving. Therefore,
the states $\left\{ \left(M_{l}\otimes\mathbb{I}\otimes\mathbb{I}\right)\left|\psi_{k}\right\rangle \right\} $,
$k=1,\dots,18$ must be orthogonal to each other. Setting the inner
products of the post measurement states equal to zero, we solve for
the matrix elements $e_{ij}$ as shown in the table below: 

\begin{equation}
\begin{array}{|c|c|}
\hline \langle\psi_{1}|\pi_{l}\otimes\mathbb{I}\otimes\mathbb{I}|\psi_{7}\rangle=0 & e_{12}=0\\
\langle\psi_{7}|\pi_{l}\otimes\mathbb{I}\otimes\mathbb{I}|\psi_{1}\rangle=0 & e_{21}=0\\
\hline \langle\psi_{3}|\pi_{l}\otimes\mathbb{I}\otimes\mathbb{I}|\psi_{9}\rangle=0 & e_{13}=0\\
\langle\psi_{9}|\pi_{l}\otimes\mathbb{I}\otimes\mathbb{I}|\psi_{3}\rangle=0 & e_{31}=0\\
\hline \langle\psi_{7}|\pi_{l}\otimes\mathbb{I}\otimes\mathbb{I}|\psi_{9}\rangle=0 & e_{23}=0\\
\langle\psi_{9}|\pi_{l}\otimes\mathbb{I}\otimes\mathbb{I}|\psi_{7}\rangle=0 & e_{32}=0\\
\hline \langle\psi_{5}|\pi_{l}\otimes\mathbb{I}\otimes\mathbb{I}|\psi_{11}\rangle=0 & e_{14}=0\\
\langle\psi_{11}|\pi_{l}\otimes\mathbb{I}\otimes\mathbb{I}|\psi_{5}\rangle=0 & e_{41}=0\\
\hline \langle\psi_{7}|\pi_{l}\otimes\mathbb{I}\otimes\mathbb{I}|\psi_{11}\rangle=0 & e_{24}=0\\
\langle\psi_{11}|\pi_{l}\otimes\mathbb{I}\otimes\mathbb{I}|\psi_{7}\rangle=0 & e_{42}=0\\
\hline \langle\psi_{9}|\pi_{l}\otimes\mathbb{I}\otimes\mathbb{I}\psi_{11}\rangle=0 & e_{34}=0\\
\langle\psi_{11}|\pi_{l}\otimes\mathbb{I}\otimes\mathbb{I}|\psi_{9}\rangle=0 & e_{43}=0\\
\hline \langle\psi_{13}|\pi_{l}\otimes\mathbb{I}\otimes\mathbb{I}|\psi_{14}\rangle=0 & e_{11}=e_{22}\\
\langle\psi_{15}|\pi_{l}\otimes\mathbb{I}\otimes\mathbb{I}|\psi_{16}\rangle=0 & e_{11}=e_{33}\\
\langle\psi_{17}|\pi_{l}\otimes\mathbb{I}\otimes\mathbb{I}|\psi_{18}\rangle=0 & e_{11}=e_{44}
\\\hline \end{array}
\end{equation}
We see that $\pi_{l}$ must be proportional to the identity. As the
argument applies to all outcomes, all the POVM elements must be proportional
to the identity, and therefore Alice cannot go first, and from the
symmetry of the states, neither can Bob, nor Charlie. Thus $S_{1}$
is locally irreducible. Since $S_{1}$ is a subset of $S$ the argument
holds for $S$ as well, and therefore, $S$ is locally irreducible. 

We now prove that $S$ is strongly nonlocal following the steps in
the proof involving the states (\ref{eq:COPB3X3X3}). Consider the
bipartition $A\vert BC$. We have already shown that Alice cannot
go first when all of them are separated (as $S$ is irreducible).
But for Alice, the situation doesn't change because she is still separated
from $BC$. Therefore, even in the configuration $A\vert BC$ Alice
cannot begin a local protocol by performing some local NOPM on her
subsystem. 
\end{spacing}

We now come to the subsystem $BC$. First we rewrite the states in
$S_{i}$, $i=1,2,3$ to reflect this fact. For clarity, we denote
computational basis states of $\mathcal{H}_{BC}$ in the following
way: $\left|1i\right\rangle \rightarrow\left|{\bf i}\right\rangle ;\left|2i\right\rangle \rightarrow\left|{\bf i+4}\right\rangle ;\left|3i\right\rangle \rightarrow\left|{\bf i+8}\right\rangle ;\left|4i\right\rangle \rightarrow\left|{\bf i+12}\right\rangle $
and rewrite the states:

\begin{spacing}{1.2}
\begin{equation}
S_{1}\left(A\vert BC\right)=\begin{array}{|ccc|}
\hline |1\rangle|{\bf 5\pm{\bf 6\rangle,}} & |2\rangle|{\bf 1\pm{\bf 5\rangle,}} & |1\pm2\rangle|{\bf 2\rangle,}\\
|1\rangle|{\bf 9\pm{\bf 11\rangle,}} & |3\rangle|{\bf 1\pm{\bf 9\rangle,}} & |1\pm3\rangle|{\bf 3\rangle,}\\
|1\rangle|{\bf 13\pm{\bf 16\rangle,}} & |4\rangle|{\bf 1\pm{\bf 13\rangle,}} & |1\pm4\rangle|{\bf 4\rangle.}
\\\hline \end{array}
\end{equation}
\begin{equation}
S_{2}\left(A|BC\right)=\begin{array}{|ccc|}
\hline |2\rangle|{\bf 9\pm{\bf 10\rangle,}} & |3\rangle|{\bf 2\pm{\bf 6\rangle,}} & |1\pm2\rangle|{\bf 7\rangle,}\\
|2\rangle|{\bf 13\pm{\bf 14\rangle,}} & |4\rangle|{\bf 2\pm{\bf 6\rangle,}} & |1\pm2\rangle|{\bf 8\rangle,}\\
|3\rangle|{\bf 13\pm{\bf 15\rangle,}} & |4\rangle|{\bf 3\pm{\bf 11\rangle,}} & |1\pm3\rangle|{\bf 12\rangle,}\\
\hline |4\rangle|{\bf 9\pm{\bf 12\rangle,}} & |3\rangle|{\bf 4\pm{\bf 16\rangle,}} & |1\pm4\rangle|{\bf 15\rangle,}\\
|4\rangle|{\bf 5\pm{\bf 8\rangle,}} & |2\rangle|{\bf 4\pm{\bf 16\rangle,}} & |1\pm4\rangle|{\bf 14\rangle,}\\
|3\rangle|{\bf 5\pm{\bf 7\rangle,}} & |2\rangle|{\bf 3\pm{\bf 11\rangle,}} & |1\pm3\rangle|{\bf 10\rangle.}
\\\hline \end{array}
\end{equation}
\begin{equation}
S_{3}\left(A\vert BC\right)=\left\{ \begin{array}{ccccc}
|2\rangle|{\bf 12\rangle,} & |3\rangle|{\bf 14\rangle,} & |4\rangle|{\bf 7\rangle,} & |2\rangle|{\bf 15\rangle,} & |4\rangle|{\bf 10\rangle,}\\
|3\rangle|{\bf 8\rangle,} & |1\rangle|{\bf 1\rangle,} & |2\rangle|{\bf 6\rangle,} & |3\rangle|{\bf 11\rangle,} & |4\rangle|{\bf 16\rangle.}
\end{array}\right\} 
\end{equation}

\end{spacing}

For any general measurement on $BC$ described by a POVM $\left\{ \pi_{\alpha}\right\} $,
each element $\pi_{\alpha}$ can be represented by a $16\times16$
matrix with the elements denoted by $a_{i,j}$, $i,j=1,\dots,16$.
We proceed exactly the same way as in the previous proofs. The table
below shows that all the diagonal elements are equal.

\begin{spacing}{1.2}
{\small{}
\begin{equation}
\begin{array}{|c|c||c|c|}
\hline \mbox{states} & \mbox{elements} & \mbox{states} & \mbox{elements}\\
\hline\hline |2\rangle|{\bf 1}+{\bf 5}\rangle,|2\rangle|{\bf 1}-{\bf 5}\rangle & a_{1,1}=a_{5,5} & |3\rangle|{\bf 1}+{\bf 9}\rangle,|3\rangle|{\bf 1}-{\bf 9}\rangle & a_{1,1}=a_{9,9}\\
\hline |4\rangle|{\bf 1}+{\bf 13}\rangle,|4\rangle|{\bf 1}-{\bf 13}\rangle & a_{1,1}=a_{13,13} & |3\rangle|{\bf 2}+{\bf 6}\rangle,|3\rangle|{\bf 2}-{\bf 6}\rangle & a_{2,2}=a_{6,6}\\
\hline |4\rangle|{\bf 3}+{\bf 11}\rangle,|4\rangle|{\bf 3}-{\bf 11}\rangle & a_{3,3}=a_{11,11} & |3\rangle|{\bf 4}+{\bf 16}\rangle,|3\rangle|{\bf 4}-{\bf 16}\rangle & a_{4,4}=a_{16,16}\\
\hline |1\rangle|{\bf 5}+{\bf 6}\rangle,|1\rangle|{\bf 5}-{\bf 6}\rangle & a_{5,5}=a_{6,6} & |1\rangle|{\bf 9}+{\bf 11}\rangle,|1\rangle|{\bf 9}-{\bf 11}\rangle & a_{9,9}=a_{11,11}\\
\hline |1\rangle|{\bf 13}+{\bf 16}\rangle,|1\rangle|{\bf 13}-{\bf 16}\rangle & a_{13,13}=a_{16,16} & |4\rangle|{\bf 5}+{\bf 8}\rangle,|4\rangle|{\bf 5}-{\bf 8}\rangle & a_{5,5}=a_{8,8}\\
\hline |3\rangle|{\bf 5}+{\bf 7}\rangle,|3\rangle|{\bf 5}-{\bf 7}\rangle & a_{5,5}=a_{7,7} & |2\rangle|{\bf 9}+{\bf 10}\rangle,|2\rangle|{\bf 9}-{\bf 10}\rangle & a_{9,9}=a_{10,10}\\
\hline |4\rangle|{\bf 9}+{\bf 12}\rangle,|4\rangle|{\bf 9}-{\bf 12}\rangle & a_{9,9}=a_{12,12} & |2\rangle|{\bf 13}+{\bf 14}\rangle,|2\rangle|{\bf 13}-{\bf 14}\rangle & a_{13,13}=a_{14,14}\\
\hline |3\rangle|{\bf 13}+{\bf 15}\rangle,|3\rangle|{\bf 13}-{\bf 15}\rangle & a_{13,13}=a_{15,15} & |2\rangle|{\bf 3}+{\bf 11}\rangle,|2\rangle|{\bf 3}-{\bf 11}\rangle & a_{3,3}=a_{11,11}
\\\hline \end{array}
\end{equation}
}Tables III and IV show that all the off-diagonal elements are zero.
The tables should be read according to the serial numbers (that's
how the matrix elements were evaluated using the orthogonality-preserving
conditions). 
\end{spacing}

Therefore we have shown that $\pi_{\alpha}$ must be proportional
to the identity. As this argument holds for all possible outcomes,
all POVM elements must also be proportional to the identity and hence,
trivial. Therefore, Bob and Charlie cannot go first. This shows that
the set is locally irreducible in the bipartition $A\vert BC$ and
from the symmetry of the states it must also be locally irreducible
in the other two bipartitions. Hence, $S$ is strongly nonlocal. 

\pagebreak

\begin{spacing}{1.2}
\begin{table}[h]
\protect\caption{Off-diagonal terms}

\centering %
\begin{tabular}{|c|c|c||c|c|c|}
\hline 
sl. no.  & states  & elements  & sl. no.  & states  & elements \tabularnewline
\hline 
\hline 
(1)  & $|1+2\rangle|{\bf 2}\rangle$, $|1+3\rangle|{\bf 3}\rangle$  & $a_{2,3}$ = $a_{3,2}$ = 0  & (2)  & $|1+2\rangle|{\bf 2}\rangle$, $|1+4\rangle|{\bf 4}\rangle$  & $a_{2,4}$ = $a_{4,2}$ = 0\tabularnewline
\hline 
(3)  & $|1+2\rangle|{\bf 2}\rangle$, $|1+2\rangle|{\bf 7}\rangle$  & $a_{2,7}$ = $a_{7,2}$ = 0  & (4)  & $|1+2\rangle|{\bf 2}\rangle$, $|1+2\rangle|{\bf 8}\rangle$  & $a_{2,8}$ = $a_{8,2}$ = 0 \tabularnewline
\hline 
(5)  & $|1+2\rangle|{\bf 2}\rangle$, $|1+3\rangle|{\bf 12}\rangle$  & $a_{2,12}$ = $a_{12,2}$ = 0  & (6)  & $|1+2\rangle|{\bf 2}\rangle$, $|1+4\rangle|{\bf 15}\rangle$  & $a_{2,15}$ = $a_{15,2}$ = 0 \tabularnewline
\hline 
(7)  & $|1+2\rangle|{\bf 2}\rangle$, $|1+4\rangle|{\bf 14}\rangle$  & $a_{2,14}$ = $a_{14,2}$ = 0  & (8)  & $|1+2\rangle|{\bf 2}\rangle$, $|1+3\rangle|{\bf 10}\rangle$  & $a_{2,10}$ = $a_{10,2}$ = 0 \tabularnewline
\hline 
(9)  & $|1+2\rangle|{\bf 2}\rangle$, $|1\rangle|{\bf 1}\rangle$  & $a_{2,1}$ = $a_{1,2}$ = 0  & (10)  & $|1+3\rangle|{\bf 3}\rangle$, $|1+4\rangle|{\bf 4}\rangle$  & $a_{3,4}$ = $a_{4,3}$ = 0\tabularnewline
\hline 
(11)  & $|1+3\rangle|{\bf 3}\rangle$, $|1+2\rangle|{\bf 7}\rangle$  & $a_{3,7}$ = $a_{7,3}$ = 0  & (12)  & $|1+3\rangle|{\bf 3}\rangle$, $|1+2\rangle|{\bf 8}\rangle$  & $a_{3,8}$ = $a_{8,3}$ = 0 \tabularnewline
\hline 
(13)  & $|1+3\rangle|{\bf 3}\rangle$, $|1+3\rangle|{\bf 12}\rangle$  & $a_{3,12}$ = $a_{12,3}$ = 0  & (14)  & $|1+3\rangle|{\bf 3}\rangle$, $|1+4\rangle|{\bf 15}\rangle$  & $a_{3,15}$ = $a_{15,3}$ = 0 \tabularnewline
\hline 
(15)  & $|1+3\rangle|{\bf 3}\rangle$, $|1+4\rangle|{\bf 14}\rangle$  & $a_{3,14}$ = $a_{14,3}$ = 0  & (16)  & $|1+3\rangle|{\bf 3}\rangle$, $|1+3\rangle|{\bf 10}\rangle$  & $a_{3,10}$ = $a_{10,3}$ = 0 \tabularnewline
\hline 
(17)  & $|1+3\rangle|{\bf 3}\rangle$, $|1\rangle|{\bf 1}\rangle$  & $a_{3,1}$ = $a_{1,3}$ = 0  & (18)  & $|1+4\rangle|{\bf 4}\rangle$, $|1+2\rangle|{\bf 7}\rangle$  & $a_{4,7}$ = $a_{7,4}$ = 0 \tabularnewline
\hline 
(19)  & $|1+4\rangle|{\bf 4}\rangle$, $|1+2\rangle|{\bf 8}\rangle$  & $a_{4,8}$ = $a_{8,4}$ = 0  & (20)  & $|1+4\rangle|{\bf 4}\rangle$, $|1+3\rangle|{\bf 12}\rangle$  & $a_{4,12}$ = $a_{12,4}$ = 0 \tabularnewline
\hline 
(21)  & $|1+4\rangle|{\bf 4}\rangle$, $|1+4\rangle|{\bf 15}\rangle$  & $a_{4,15}$ = $a_{15,4}$ = 0  & (22)  & $|1+4\rangle|{\bf 4}\rangle$, $|1+4\rangle|{\bf 14}\rangle$  & $a_{4,14}$ = $a_{14,4}$ = 0 \tabularnewline
\hline 
(23)  & $|1+4\rangle|{\bf 4}\rangle$, $|1+3\rangle|{\bf 10}\rangle$  & $a_{4,10}$ = $a_{10,4}$ = 0  & (24)  & $|1+4\rangle|{\bf 4}\rangle$, $|1\rangle|{\bf 1}\rangle$  & $a_{4,1}$ = $a_{1,4}$ = 0 \tabularnewline
\hline 
(25)  & $|1+2\rangle|{\bf 7}\rangle$, $|1+2\rangle|{\bf 8}\rangle$  & $a_{7,8}$ = $a_{8,7}$ = 0  & (26)  & $|1+2\rangle|{\bf 7}\rangle$, $|1+3\rangle|{\bf 12}\rangle$  & $a_{7,12}$ = $a_{12,7}$ = 0 \tabularnewline
\hline 
(27)  & $|1+2\rangle|{\bf 7}\rangle$, $|1+4\rangle|{\bf 15}\rangle$  & $a_{7,15}$ = $a_{15,7}$ = 0  & (28)  & $|1+2\rangle|{\bf 7}\rangle$, $|1+4\rangle|{\bf 14}\rangle$  & $a_{7,14}$ = $a_{14,7}$ = 0 \tabularnewline
\hline 
(29)  & $|1+2\rangle|{\bf 7}\rangle$, $|1+3\rangle|{\bf 10}\rangle$  & $a_{7,10}$ = $a_{10,7}$ = 0  & (30)  & $|1+2\rangle|{\bf 7}\rangle$, $|1\rangle|{\bf 1}\rangle$  & $a_{7,1}$ = $a_{1,7}$ = 0 \tabularnewline
\hline 
(31)  & $|1+2\rangle|{\bf 8}\rangle$, $|1+3\rangle|{\bf 12}\rangle$  & $a_{8,12}$ = $a_{12,8}$ = 0  & (32)  & $|1+2\rangle|{\bf 8}\rangle$, $|1+4\rangle|{\bf 15}\rangle$  & $a_{8,15}$ = $a_{15,8}$ = 0 \tabularnewline
\hline 
(33)  & $|1+2\rangle|{\bf 8}\rangle$, $|1+4\rangle|{\bf 14}\rangle$  & $a_{8,14}$ = $a_{14,8}$ = 0  & (34)  & $|1+2\rangle|{\bf 8}\rangle$, $|1+3\rangle|{\bf 10}\rangle$  & $a_{8,10}$ = $a_{10,8}$ = 0 \tabularnewline
\hline 
(35)  & $|1+2\rangle|{\bf 8}\rangle$, $|1\rangle|{\bf 1}\rangle$  & $a_{8,1}$ = $a_{1,8}$ = 0  & (36)  & $|1+3\rangle|{\bf 12}\rangle$, $|1+4\rangle|{\bf 15}\rangle$  & $a_{12,15}$ = $a_{15,12}$ = 0 \tabularnewline
\hline 
(37)  & $|1+3\rangle|{\bf 12}\rangle$, $|1+4\rangle|{\bf 14}\rangle$  & $a_{12,14}$ = $a_{14,12}$ = 0  & (38)  & $|1+3\rangle|{\bf 12}\rangle$, $|1+3\rangle|{\bf 10}\rangle$  & $a_{12,10}$ = $a_{10,12}$ = 0 \tabularnewline
\hline 
(39)  & $|1+3\rangle|{\bf 12}\rangle$, $|1\rangle|{\bf 1}\rangle$  & $a_{12,1}$ = $a_{1,12}$ = 0  & (40)  & $|1+4\rangle|{\bf 15}\rangle$, $|1+4\rangle|{\bf 14}\rangle$  & $a_{15,14}$ = $a_{14,15}$ = 0 \tabularnewline
\hline 
(41)  & $|1+4\rangle|{\bf 15}\rangle$, $|1+3\rangle|{\bf 10}\rangle$  & $a_{15,10}$ = $a_{10,15}$ = 0  & (42)  & $|1+4\rangle|{\bf 15}\rangle$, $|1\rangle|{\bf 1}\rangle$  & $a_{15,1}$ = $a_{1,15}$ = 0 \tabularnewline
\hline 
(43)  & $|1+4\rangle|{\bf 14}\rangle$, $|1+3\rangle|{\bf 10}\rangle$  & $a_{14,10}$ = $a_{10,14}$ = 0  & (44)  & $|1+4\rangle|{\bf 14}\rangle$, $|1\rangle|{\bf 1}\rangle$  & $a_{14,1}$ = $a_{1,14}$ = 0 \tabularnewline
\hline 
(45)  & $|1+3\rangle|{\bf 10}\rangle$, $|1\rangle|{\bf 1}\rangle$  & $a_{10,1}$ = $a_{1,10}$ = 0  & (46)  & $|1+2\rangle|{\bf 2}\rangle$, $|2\rangle|{\bf {3+11}}\rangle$  & $a_{2,11}$ = $a_{11,2}$ = 0 \tabularnewline
\hline 
(47)  & $|1+2\rangle|{\bf 2}\rangle$, $|2\rangle|{\bf {4+16}}\rangle$  & $a_{2,16}$ = $a_{16,2}$ = 0  & (48)  & $|1+2\rangle|{\bf 2}\rangle$, $|2\rangle|{\bf {9+10}}\rangle$  & $a_{2,9}$ = $a_{9,2}$ = 0 \tabularnewline
\hline 
(49)  & $|1+2\rangle|{\bf 2}\rangle$, $|2\rangle|{\bf {6}}\rangle$  & $a_{2,6}$ = $a_{6,2}$ = 0  & (50)  & $|1+2\rangle|{\bf 2}\rangle$, $|2\rangle|{\bf {13+14}}\rangle$  & $a_{2,13}$ = $a_{13,2}$ = 0 \tabularnewline
\hline 
(51)  & $|1+2\rangle|{\bf 2}\rangle$, $|1\rangle|{\bf {5+6}}\rangle$  & $a_{2,5}$ = $a_{5,2}$ = 0  & (52)  & $|1+2\rangle|{\bf 7}\rangle$, $|2\rangle|{\bf {3+11}}\rangle$  & $a_{7,11}$ = $a_{11,7}$ = 0 \tabularnewline
\hline 
(53)  & $|1+2\rangle|{\bf 7}\rangle$, $|2\rangle|{\bf {4+16}}\rangle$  & $a_{7,16}$ = $a_{16,7}$ = 0  & (54)  & $|1+2\rangle|{\bf 7}\rangle$, $|2\rangle|{\bf {9+10}}\rangle$  & $a_{7,9}$ = $a_{9,7}$ = 0 \tabularnewline
\hline 
(55)  & $|1+2\rangle|{\bf 7}\rangle$, $|2\rangle|{\bf {6}}\rangle$  & $a_{7,6}$ = $a_{6,7}$ = 0  & (56)  & $|1+2\rangle|{\bf 7}\rangle$, $|2\rangle|{\bf {13+14}}\rangle$  & $a_{7,13}$ = $a_{13,7}$ = 0 \tabularnewline
\hline 
(57)  & $|1+2\rangle|{\bf 7}\rangle$, $|1\rangle|{\bf {5+6}}\rangle$  & $a_{7,5}$ = $a_{5,7}$ = 0  & (58)  & $|1+2\rangle|{\bf 8}\rangle$, $|2\rangle|{\bf {3+11}}\rangle$  & $a_{8,11}$ = $a_{11,8}$ = 0 \tabularnewline
\hline 
(59)  & $|1+2\rangle|{\bf 8}\rangle$, $|2\rangle|{\bf {4+16}}\rangle$  & $a_{8,16}$ = $a_{16,8}$ = 0  & (60)  & $|1+2\rangle|{\bf 8}\rangle$, $|2\rangle|{\bf {9+10}}\rangle$  & $a_{8,9}$ = $a_{9,8}$ = 0 \tabularnewline
\hline 
\end{tabular}
\end{table}

\end{spacing}

\pagebreak

\begin{table}[h]
\protect\caption{Off-diagonal terms}

\centering %
\begin{tabular}{|c|c|c||c|c|c|}
\hline 
sl. no.  & states  & elements  & sl. no.  & states  & elements \tabularnewline
\hline 
\hline 
(61)  & $|1+2\rangle|{\bf 8}\rangle$, $|2\rangle|{\bf {6}}\rangle$  & $a_{8,6}$ = $a_{6,8}$ = 0  & (62)  & $|1+2\rangle|{\bf 8}\rangle$, $|2\rangle|{\bf {13+14}}\rangle$  & $a_{8,13}$ = $a_{13,8}$ = 0 \tabularnewline
\hline 
(63)  & $|1+2\rangle|{\bf 8}\rangle$, $|1\rangle|{\bf {5+6}}\rangle$  & $a_{8,5}$ = $a_{5,8}$ = 0  & (64)  & $|1+3\rangle|{\bf {3}}\rangle$, $|3\rangle|{\bf {11}}\rangle$  & $a_{3,11}$ = $a_{11,3}$ = 0 \tabularnewline
\hline 
(65)  & $|1+3\rangle|{\bf {3}}\rangle$, $|3\rangle|{\bf {4+16}}\rangle$  & $a_{3,16}$ = $a_{16,3}$ = 0  & (66)  & $|1+3\rangle|{\bf {3}}\rangle$, $|3\rangle|{\bf {1+9}}\rangle$  & $a_{3,9}$ = $a_{9,3}$ = 0 \tabularnewline
\hline 
(67)  & $|1+3\rangle|{\bf {3}}\rangle$, $|3\rangle|{\bf {2+6}}\rangle$  & $a_{3,6}$ = $a_{6,3}$ = 0  & (68)  & $|1+3\rangle|{\bf {3}}\rangle$, $|3\rangle|{\bf {13+15}}\rangle$  & $a_{3,13}$ = $a_{13,3}$ = 0 \tabularnewline
\hline 
(69)  & $|1+3\rangle|{\bf {3}}\rangle$, $|3\rangle|{\bf {5+7}}\rangle$  & $a_{3,5}$ = $a_{5,3}$ = 0  & (70)  & $|1+3\rangle|{\bf {12}}\rangle$, $|3\rangle|{\bf {11}}\rangle$  & $a_{12,11}$ = $a_{11,12}$ = 0 \tabularnewline
\hline 
(71)  & $|1+3\rangle|{\bf {12}}\rangle$, $|3\rangle|{\bf {4+16}}\rangle$  & $a_{12,16}$ = $a_{16,12}$ = 0  & (72)  & $|1+3\rangle|{\bf {12}}\rangle$, $|3\rangle|{\bf {1+9}}\rangle$  & $a_{12,9}$ = $a_{9,12}$ = 0 \tabularnewline
\hline 
(73)  & $|1+3\rangle|{\bf {12}}\rangle$, $|3\rangle|{\bf {2+6}}\rangle$  & $a_{12,6}$ = $a_{6,12}$ = 0  & (74)  & $|1+3\rangle|{\bf {12}}\rangle$, $|3\rangle|{\bf {13+15}}\rangle$  & $a_{12,13}$ = $a_{13,12}$ = 0 \tabularnewline
\hline 
(75)  & $|1+3\rangle|{\bf {12}}\rangle$, $|3\rangle|{\bf {5+7}}\rangle$  & $a_{12,5}$ = $a_{5,12}$ = 0  & (76)  & $|1+3\rangle|{\bf {10}}\rangle$, $|3\rangle|{\bf {11}}\rangle$  & $a_{10,11}$ = $a_{11,10}$ = 0 \tabularnewline
\hline 
(77)  & $|1+3\rangle|{\bf {10}}\rangle$, $|3\rangle|{\bf {4+16}}\rangle$  & $a_{10,16}$ = $a_{16,10}$ = 0  & (78)  & $|1+3\rangle|{\bf {10}}\rangle$, $|3\rangle|{\bf {1+9}}\rangle$  & $a_{10,9}$ = $a_{9,10}$ = 0 \tabularnewline
\hline 
(79)  & $|1+3\rangle|{\bf {10}}\rangle$, $|3\rangle|{\bf {2+6}}\rangle$  & $a_{10,6}$ = $a_{6,10}$ = 0  & (80)  & $|1+3\rangle|{\bf {10}}\rangle$, $|3\rangle|{\bf {13+15}}\rangle$  & $a_{10,13}$ = $a_{13,10}$ = 0 \tabularnewline
\hline 
(81)  & $|1+3\rangle|{\bf {10}}\rangle$, $|3\rangle|{\bf {5+7}}\rangle$  & $a_{10,5}$ = $a_{5,10}$ = 0  & (82)  & $|1+4\rangle|{\bf {4}}\rangle$, $|4\rangle|{\bf {3+11}}\rangle$  & $a_{4,11}$ = $a_{11,4}$ = 0 \tabularnewline
\hline 
(83)  & $|1+4\rangle|{\bf {4}}\rangle$, $|4\rangle|{\bf {16}}\rangle$  & $a_{4,16}$ = $a_{16,4}$ = 0  & (84)  & $|1+4\rangle|{\bf {4}}\rangle$, $|4\rangle|{\bf {9+12}}\rangle$  & $a_{4,9}$ = $a_{9,4}$ = 0 \tabularnewline
\hline 
(85)  & $|1+4\rangle|{\bf {4}}\rangle$, $|4\rangle|{\bf {5+8}}\rangle$  & $a_{4,5}$ = $a_{5,4}$ = 0  & (86)  & $|1+4\rangle|{\bf {4}}\rangle$, $|4\rangle|{\bf {2+6}}\rangle$  & $a_{4,6}$ = $a_{6,4}$ = 0 \tabularnewline
\hline 
(87)  & $|1+4\rangle|{\bf {4}}\rangle$, $|4\rangle|{\bf {1+13}}\rangle$  & $a_{4,13}$ = $a_{13,4}$ = 0  & (88)  & $|1+4\rangle|{\bf {15}}\rangle$, $|4\rangle|{\bf {3+11}}\rangle$  & $a_{15,11}$ = $a_{11,15}$ = 0 \tabularnewline
\hline 
(89)  & $|1+4\rangle|{\bf {15}}\rangle$, $|4\rangle|{\bf {16}}\rangle$  & $a_{15,16}$ = $a_{16,15}$ = 0  & (90)  & $|1+4\rangle|{\bf {15}}\rangle$, $|4\rangle|{\bf {9+12}}\rangle$  & $a_{15,9}$ = $a_{9,15}$ = 0 \tabularnewline
\hline 
(91)  & $|1+4\rangle|{\bf {15}}\rangle$, $|4\rangle|{\bf {5+8}}\rangle$  & $a_{15,5}$ = $a_{5,15}$ = 0  & (92)  & $|1+4\rangle|{\bf {15}}\rangle$, $|4\rangle|{\bf {2+6}}\rangle$  & $a_{15,6}$ = $a_{6,15}$ = 0 \tabularnewline
\hline 
(93)  & $|1+4\rangle|{\bf {15}}\rangle$, $|4\rangle|{\bf {1+13}}\rangle$  & $a_{15,13}$ = $a_{13,15}$ = 0  & (94)  & $|1+4\rangle|{\bf {14}}\rangle$, $|4\rangle|{\bf {3+11}}\rangle$  & $a_{14,11}$ = $a_{11,14}$ = 0 \tabularnewline
\hline 
(95)  & $|1+4\rangle|{\bf {14}}\rangle$, $|4\rangle|{\bf {16}}\rangle$  & $a_{14,16}$ = $a_{16,14}$ = 0  & (96)  & $|1+4\rangle|{\bf {14}}\rangle$, $|4\rangle|{\bf {9+12}}\rangle$  & $a_{14,9}$ = $a_{9,14}$ = 0 \tabularnewline
\hline 
(97)  & $|1+4\rangle|{\bf {14}}\rangle$, $|4\rangle|{\bf {5+8}}\rangle$  & $a_{14,5}$ = $a_{5,14}$ = 0  & (98)  & $|1+4\rangle|{\bf {14}}\rangle$, $|4\rangle|{\bf {2+6}}\rangle$  & $a_{14,6}$ = $a_{6,14}$ = 0 \tabularnewline
\hline 
(99)  & $|1+4\rangle|{\bf {14}}\rangle$, $|4\rangle|{\bf {1+13}}\rangle$  & $a_{14,13}$ = $a_{13,14}$ = 0  & (100)  & $|3\rangle|{\bf {5+7}}\rangle$, $|3\rangle|{\bf {11}}\rangle$  & $a_{5,11}$ = $a_{11,5}$ = 0 \tabularnewline
\hline 
(101)  & $|3\rangle|{\bf {5+7}}\rangle$, $|3\rangle|{\bf {4+16}}\rangle$  & $a_{5,16}$ = $a_{16,5}$ = 0  & (102)  & $|3\rangle|{\bf {5+7}}\rangle$, $|3\rangle|{\bf {2+6}}\rangle$  & $a_{5,6}$ = $a_{6,5}$ = 0 \tabularnewline
\hline 
(103)  & $|3\rangle|{\bf {5+7}}\rangle$, $|3\rangle|{\bf {13+15}}\rangle$  & $a_{5,13}$ = $a_{13,5}$ = 0  & (104)  & $|4\rangle|{\bf {5+8}}\rangle$, $|4\rangle|{\bf {9+12}}\rangle$  & $a_{5,9}$ = $a_{9,5}$ = 0 \tabularnewline
\hline 
(105)  & $|3\rangle|{\bf {2+6}}\rangle$, $|3\rangle|{\bf {4+16}}\rangle$  & $a_{6,16}$ = $a_{16,6}$ = 0  & (106)  & $|3\rangle|{\bf {2+6}}\rangle$, $|3\rangle|{\bf {11}}\rangle$  & $a_{6,11}$ = $a_{11,6}$ = 0 \tabularnewline
\hline 
(107)  & $|3\rangle|{\bf {2+6}}\rangle$, $|3\rangle|{\bf {13+15}}\rangle$  & $a_{6,13}$ = $a_{13,6}$ = 0  & (108)  & $|4\rangle|{\bf {2+6}}\rangle$, $|4\rangle|{\bf {9+12}}\rangle$  & $a_{6,9}$ = $a_{9,6}$ = 0 \tabularnewline
\hline 
(109)  & $|4\rangle|{\bf {9+12}}\rangle$, $|4\rangle|{\bf {3+11}}\rangle$  & $a_{9,11}$ = $a_{11,9}$ = 0  & (110)  & $|4\rangle|{\bf {9+12}}\rangle$, $|4\rangle|{\bf {16}}\rangle$  & $a_{9,16}$ = $a_{16,9}$ = 0 \tabularnewline
\hline 
(111)  & $|2\rangle|{\bf {9+10}}\rangle$, $|2\rangle|{\bf {13+14}}\rangle$  & $a_{9,13}$ = $a_{13,9}$ = 0  & (112)  & $|4\rangle|{\bf {3+11}}\rangle$, $|4\rangle|{\bf {16}}\rangle$  & $a_{11,16}$ = $a_{16,11}$ = 0 \tabularnewline
\hline 
(113)  & $|2\rangle|{\bf {3+11}}\rangle$, $|2\rangle|{\bf {13+14}}\rangle$  & $a_{11,13}$ = $a_{13,11}$ = 0  & (114)  & $|2\rangle|{\bf {13+14}}\rangle$, $|2\rangle|{\bf {4+16}}\rangle$  & $a_{13,16}$ = $a_{16,13}$ = 0 \tabularnewline
\hline 
(115)  & $|2\rangle|{\bf {1+5}}\rangle$, $|2\rangle|{\bf {6}}\rangle$  & $a_{1,6}$ = $a_{6,1}$ = 0  & (116)  & $|2\rangle|{\bf {1+5}}\rangle$, $|2\rangle|{\bf {13+14}}\rangle$  & $a_{1,13}$ = $a_{13,1}$ = 0 \tabularnewline
\hline 
(117)  & $|2\rangle|{\bf {1+5}}\rangle$, $|2\rangle|{\bf {3+11}}\rangle$  & $a_{1,11}$ = $a_{11,1}$ = 0  & (118)  & $|2\rangle|{\bf {1+5}}\rangle$, $|2\rangle|{\bf {4+16}}\rangle$  & $a_{1,16}$ = $a_{16,1}$ = 0 \tabularnewline
\hline 
(119)  & $|2\rangle|{\bf {1+5}}\rangle$, $|2\rangle|{\bf {9+10}}\rangle$  & $a_{1,9}$ = $a_{9,1}$ = 0  & (120)  & $|3\rangle|{\bf {1+9}}\rangle$, $|3\rangle|{\bf {5+7}}\rangle$  & $a_{1,5}$ = $a_{5,1}$ = 0 \tabularnewline
\hline 
\end{tabular}
\end{table}

\medskip

\twocolumngrid

\medskip
\end{document}